\documentclass[letterpaper, numberedappendix]{emulateapj} 

\usepackage{graphicx}
\usepackage{epsfig}
\usepackage{amsmath}
\usepackage{subfigure} 
\usepackage{hyperref}
\usepackage{amssymb}
\usepackage{lscape} 
\usepackage{apjfonts}
\usepackage{latexsym}
\usepackage[usenames,dvipsnames]{color}

\newcommand{\vx}[0]{\mathbf{x}}
\newcommand{\vm}[0]{\mathbf{m}}

\newcommand{\tabref}[1]{Table \,\ref{#1}}

\newcommand{\MW}[1]{$\text{MW}{#1}$}

\DeclareMathAlphabet{\mathitbf}{OML}{cmm}{b}{it}

\shorttitle{Dissecting galaxy formation models} 

\shortauthors{F.~A. G\'omez \& C.E. Coleman-Smith et al.}

\begin{document}\title{Dissecting   galaxy   formation   models   with
  Sensitivity Analysis  -- A new  approach to constrain  the Milky Way
  formation  history}

\author{Facundo A. G\'omez}\affil{Department of Physics and Astronomy,
  Michigan State University, East Lansing, MI 48824, USA}\affil{Institute
  for Cyber-Enabled Research, Michigan State University, East Lansing, MI
  48824, USA}  
\author{Christopher E. Coleman-Smith}\affil{Department of Physics, Duke
  University, Durham, NC, 27708, USA}  
\author{Brian W. O'Shea}\affil{Department of Physics and Astronomy,
  Michigan State University, East Lansing, MI 48824,
  USA}\affil{Lyman Briggs College,
  Michigan State University, East Lansing, MI 48825, USA}
\affil{Institute for Cyber-Enabled Research, Michigan State
  University, East Lansing, MI 48824, USA}\affil{Joint
  Institute for Nuclear Astrophysics, Michigan State University, East Lansing, MI 48824, USA }
\author{Jason Tumlinson}\affil{Space Telescope Science Institute,
  Baltimore, MD, USA} 
\author{Robert. L. Wolpert}
\affil{Department of Statistical Science, Duke University, Durham, NC
  27708-0251}. 

\label{firstpage}

\begin{abstract}
  We present  an application  of a  statistical tool  known as
    Sensitivity  Analysis  to  characterize the  relationship  between
    input  parameters and  observational predictions  of semi-analytic
    models  of  galaxy  formation  coupled  to  cosmological  $N$-body
    simulations. We  show how a sensitivity analysis  can be performed
    on  our  chemo-dynamical  model,  ChemTreeN, to  characterize  and
    quantify  its  relationship  between  model input  parameters  and
    predicted  observable properties.  The result  of  this analysis
  provides the  user with information about which  parameters are most
  important  and most  likely  to  affect the  prediction  of a  given
  observable.  It can  also be used to simplify  models by identifying
  input  parameters  that  have   no  effect  on  the  outputs  (i.e.,
  observational  predictions)  of  interest.  Conversely,  sensitivity
  analysis  allows us  to identify  what model  parameters can  be most
  efficiently  constrained by  the given  observational data  set.  We
  have  applied  this  technique   to  real  observational  data  sets
  associated with the Milky Way,  such as the luminosity function of the
  dwarf satellites.   The  results from the  sensitivity analysis
    are  used to  train specific  model emulators  of  ChemTreeN, only
    involving the most relevant  input parameters.  This allowed us to
    efficiently  explore the  input parameter  space.   A statistical
  comparison of model outputs and real observables is used to obtain a
  ``best-fitting''  parameter   set.   We  consider   different  Milky
  Way-like  dark matter  halos to  account for  the dependence  of the
  best-fitting parameters  selection process on  the underlying merger
  history  of the  models.   For all  formation histories  considered,
  running ChemTreeN  with best-fitting parameters  produced luminosity
  functions  that tightly  fit their  observed  counterpart.  However,
  only one of the resulting  stellar halo models was able to reproduce
  the observed stellar halo mass within 40 kpc of the Galactic center.
  On the  basis of this analysis  it is possible  to disregard certain
  models,   and  their   corresponding  merger   histories,   as  good
  representations of the underlying merger history of the Milky Way.

\end{abstract}

\keywords{galaxies:  formation --  Galaxy: formation  -- Galaxy:  halo --
  methods: analytical -- methods: numerical -- methods: statistical}

\section{Introduction}
\label{sec:intro}

The study of galaxy formation presents many theoretical challenges.  A
huge range of physical processes come into play, and often interact in
nonlinear  ways.  Models of  galaxy formation  are rapidly  growing in
complexity to address both the  physics we believe is required as well
as  the  ever-expanding  observational  details  \citep[for  a  recent
review, see  ][]{2010PhR...495...33B}.  A recent example  of this
  observationally-driven model  evolution came as a result  of what is
  known      as       the      ``missing      satellite      problem''
  \citep{1999ApJ...524L..19M,  1999ApJ...522...82K}. The overabundance
  of dark  matter satellites in cosmological  simulations with respect
  to the  number of observed  luminous satellites in, e.g.,  the Milky
  Way  and   M31  can  be  significantly  alleviated   thanks  to  the
  suppression of star formation in  small halos that occurs during the
  epoch  of re-ionization  \citep{ 2000ApJ...539..517B,  gne}. Current
  models of galaxy formation include phenomenological prescriptions to
  treat   this  process   as  one   of  their   basic   aspects.   The
  luminosity-metallicity  relation  observed  for  Local  Group  dwarf
  galaxies  is  another example  of  complex  physical processes  that
  required the addition of new prescriptions to reproduce the available
  data  sets \citep{2003MNRAS.344.1131D}.   Both model  parameters and
  available  observational  constraints are  growing  at an  extremely
  rapid pace.

Theoretical  models   include   semi-analytic   models  such  as
ChemTreeN,              Galform,             or             Galacticus
\citep{tumlinson2,2010MNRAS.407.2017B,2012NewA...17..175B}.       These
models  use  either extended  Press-Schechter  or N-body  cosmological
simulations   to   provide   galaxy   merger  histories,   and they   apply
prescriptions  for the evolution  of the  baryonic components  of the
universe  on  top   of  this.   Similarly,  physics-rich  cosmological
simulations model  the formation of galaxies  in unprecedented detail,
with  a separate  set of  strengths  and limitations.   Both types  of
models are providing predictions  about the distribution of observable
quantities for galaxies --  particularly Milky Way-type galaxies -- in
great  detail.   Recent  examples include  metallicity  \citep{cooper,
  2011MNRAS.416.2802F,G12,2013MNRAS.432.3391T,2013arXiv1309.3609T},
stellar   chemical   abundances   \citep{2006ApJ...638..585F},   color
profiles   \citep{anto},  luminosity   and  radial   distributions  of
satellite       galaxies      \citep{2008ApJ...686..279K,      toll08,
  2013MNRAS.429.1502W},  and   the  degree  of   substructure  in  the
phase-space of the stellar halo \citep{2013arXiv1307.0008G}.

Current and upcoming  observational campaigns are providing tremendous
amounts of  data about  the Milky Way  and other galaxies.   The Sloan
Digital  Sky  Survey (SDSS), both  through  the  photometric  survey and  the
spectroscopic SEGUE  project, has  truly revolutionized our  study of
the  Milky  Way  and   its  satellites,  including  finding  many  new
ultra-faint                       dwarf                       galaxies
\citep[e.g.][]{sdss,2006ApJ...642L.137B,2007ApJ...654..897B,bell08,segue,2010ApJ...712L.103B}.
SEGUE and  the RAVE project  \citep{rave} have also  provided velocity
and metallicity information about  huge numbers of stars, allowing new
discoveries    to    be    made    \citep[e.g.][]{2008Natur.451..216C,
  2008ApJ...684..287I,    2008ApJ...673..864J,    2010ApJ...712..692C,
  2010ApJ...716....1B,2012MNRAS.423.3727G,         2012ApJ...750L..41W,
  2012sf2a.conf..121S,2013MNRAS.tmp.2356W}.   The SDSS  APOGEE project
will extend our understanding of the chemical properties of the bulge,
disk, and  halo \citep{apogee},  and the LAMOST  spectroscopic project
\citep{lamost}  will   increase  the  number  of   halo  stars  found,
supplementing the  data taken by SEGUE  and RAVE.  In  the future, the
SkyMapper   project   \citep{2012ASPC..458..409K},   the  {\it   Gaia}
satellite   \citep{2001A&A...369..339P},    and   their   accompanying
high-resolution  spectroscopic  follow-up  campaigns  \citep{  hermes,
  2012Msngr.147...25G}  will produce  even  more detailed  information
about  Milky Way stellar  populations, providing  a vastly  larger and
more  uniform sample  of high-resolution  abundance  measurements than
currently exists \citep[e.g.,][]{2013pss5.book...55F}.  In addition to
the Milky Way and its satellites, detailed observations are being made
of  the  stellar  halos  and  satellite  populations  of  other  Milky
Way-sized                                                      galaxies
\citep[e.g.,][]{2005ApJ...633..828M,2009Natur.461...66M,
  2011ApJS..195...18R,2012ApJ...760...76G,anto}.

Taken together, recent and projected  advances on all fronts in galaxy
formation suggest that we are entering an era where robust statistical
comparison  between  models and  observations  is essential.   Several
efforts are underway to develop the tools required for this enterprise
\citep{ 2009MNRAS.396..535H, 2010MNRAS.407.2017B, 2012MNRAS.421.1779L,
  G12,  2013arXiv1310.7034R, 2013arXiv1311.0047L}.   In most  of these
works, the  main goal  was the identification  of the ``best''  set of
input parameters, or a region of best fit, within which a given set of
observations  could be  successfully reproduced  by a  specific model.
However,  due  to  the  growing  complexity  of  the  models  and  the
non-linear coupling between physical processes therein, it is becoming
increasingly important to incorporate statistical tools that allow one
to  identify and  quantify the  significance of  relationships between
input  parameters  and  observable  predictions.   \textit{Sensitivity
  analysis} is  an example of  this kind of statistical  method, which
provides  a  systematic  and  quantitative way  of  understanding  the
parameters that have the most influence in a given model and, in turn,
could be most readily constrained with a given observational data set.
It can also be used to simplify models by identifying input parameters
that  have  minimal influence  on  the  available  set of  outputs  or
observables.

In  this  work we  demonstrate  the  use  of sensitivity  analysis  in
achieving   the  goals  of   both  quantitatively   and  qualitatively
understanding   the  relationships   between   input  parameters   and
observable predictions  for galaxy formation models  -- in particular,
ChemTreeN,  a  semi-analytic  model  that  has been  used  in  several
previous works \citep{tumlinson1,tumlinson2, G12,2013ApJ...773..105C}.
Such an  analysis requires  a very dense  sampling of models  within a
high-dimensional space  of input parameters.  To make  the project
computationally  feasible we supplement  ChemTreeN with  a statistical
surrogate model known  as a Gaussian process emulator.   This tool can
be used  to give predictions for  both model outputs  and an attendant
measure  of  uncertainty about  these  outputs  at  any point  in  the
parameter space,  and it is ``trained''  using a set of  galaxy evolution
models    that   span    the   required    space   \citep[][hereafter,
G12]{2010MNRAS.407.2017B,G12}. Through  the combination of sensitivity
analysis  and Gaussian  process model  emulation, we  can  rapidly and
reliably achieve our stated goals.

In addition to performing a sensitivity analysis, we apply our
statistical machinery to an observational data set obtained from the
Milky Way's satellite dwarf galaxies.  Guided by the results provided
by the sensitivity analysis, we look for constraints on our model
parameters from different observable quantities.  As was previously
shown by G12, we find that the best-fitting parameter values
strongly depend on the merger history of the model being considered.
Furthermore, we show how it is possible to constrain the formation
history of the Milky Way by contrasting the best-fitting models to an
independent set of observables.

This  paper  is   structured  as  follows.   Section~\ref{sec:methods}
briefly  describes  the  components  of our  galaxy  evolution  model,
including   the    N-body   simulations   and    the   model   itself.
Section~\ref{sec:mod_emu}   describes  the   Gaussian   process  model
emulator and our  sensitivity analysis.  Section~\ref{sec:in-out} uses
these  models and  statistical tools  to understand  the relationships
between the  input parameters and  observable predictions made  by the
models,  and   Section~\ref{sec:realmw} uses  observations  of the
  Milky Way dwarf galaxy population to show how these techniques could
  help us  to constrain the Milky  Way's formation history  as well as
  its  properties  at  $z=0$.    Finally,  we  discuss  some  of  the
limitations   of   this   work    and   summarize   our   results   in
Section~\ref{sec:conclusions}.

Throughout this study we work with  both mock and real observational data sets. From
now on  we will refer to them as mock and real observables, respectively.

\begin{deluxetable}{lcccc}
  \tabletypesize{\footnotesize} 
  \tablecaption{Main properties at $z=0$ of
    the four Dark Matter halos analyzed in this work. \label{tab:sims}}
  \tablewidth{230pt} \tablehead{\colhead{Name}&
    \colhead{$R_{200}$\tablenotemark{a}} &
    \colhead{$M_{200}$\tablenotemark{b}} & \colhead{$c$} & \colhead{$z_{\rm
        LMM}$}} \startdata
  \MW1 & 381 & 1.63 & 12.2 & 2.1 \\
  \MW2 & 378 & 1.59 & 9.2 & 3.5 \\
  \MW3 & 347 & 1.23 & 15.5 & 2.0 \\
  \MW4\tablenotemark{c} & 366 & 1.44 & 13.6 & 3.0
\enddata
\tablecomments{From  left to  right, the  columns give  the simulation
  label, the virial radius of the dark matter halo, $R_{200}$, the mass
  within $R_{200}$,  $M_{200}$, the concentration  parameter, $c$, and
  the   redshift   of  the   last   mayor   merger,  $z_{\rm  LMM}$.}
\tablenotetext{a}{Distances       are       listed       in       kpc}
\tablenotetext{b}{Masses    are    listed   in    $10^{12}~M_{\odot}$}
\tablenotetext{c}{\MW4 corresponds to  the simulation MW6 presented in
  T10}
\end{deluxetable} 

\section{Numerical methods}
\label{sec:methods}

\begin{figure*}
\centering
\hspace{-0.5cm}
\includegraphics[width=86mm,clip]{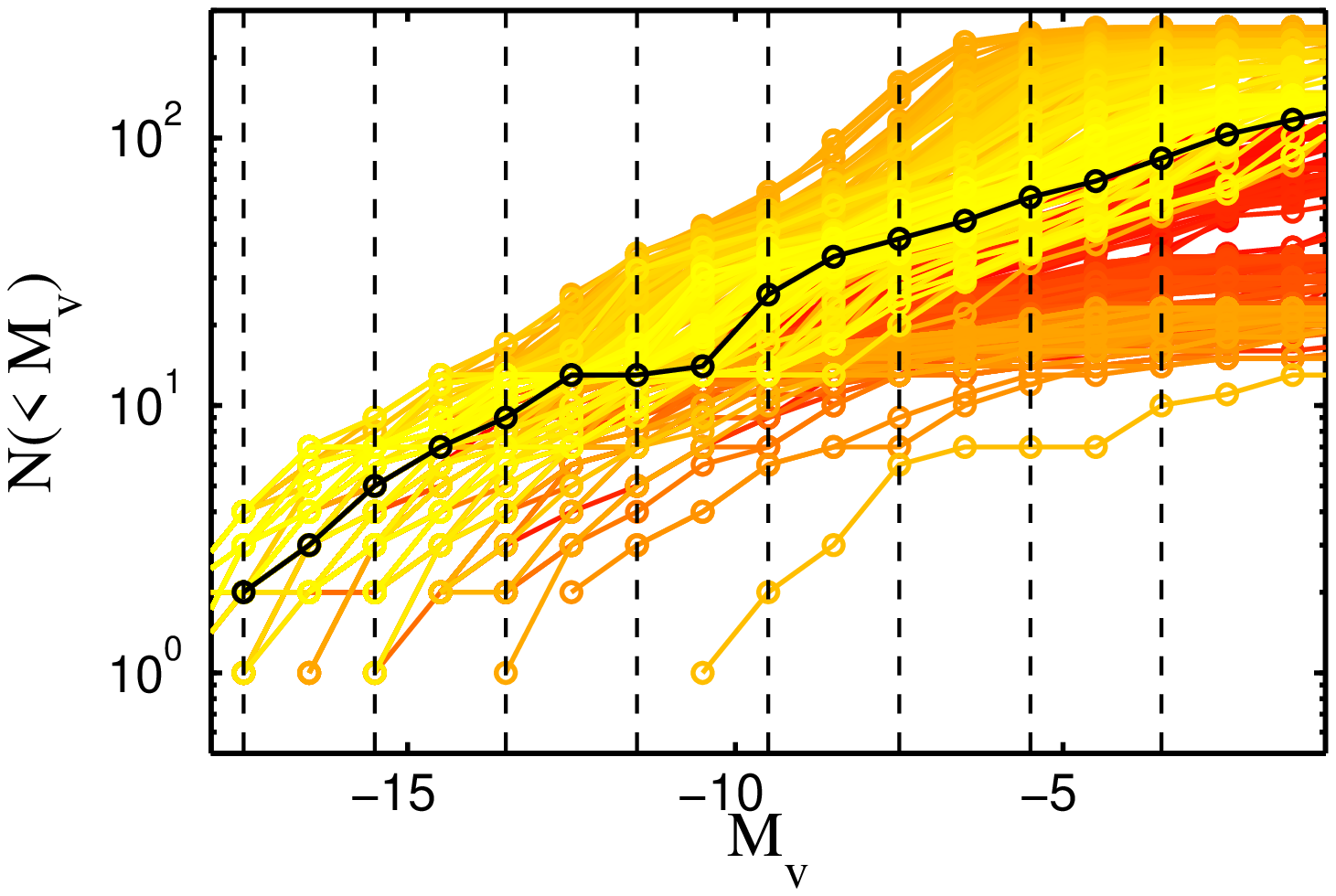}
\includegraphics[width=86mm,clip]{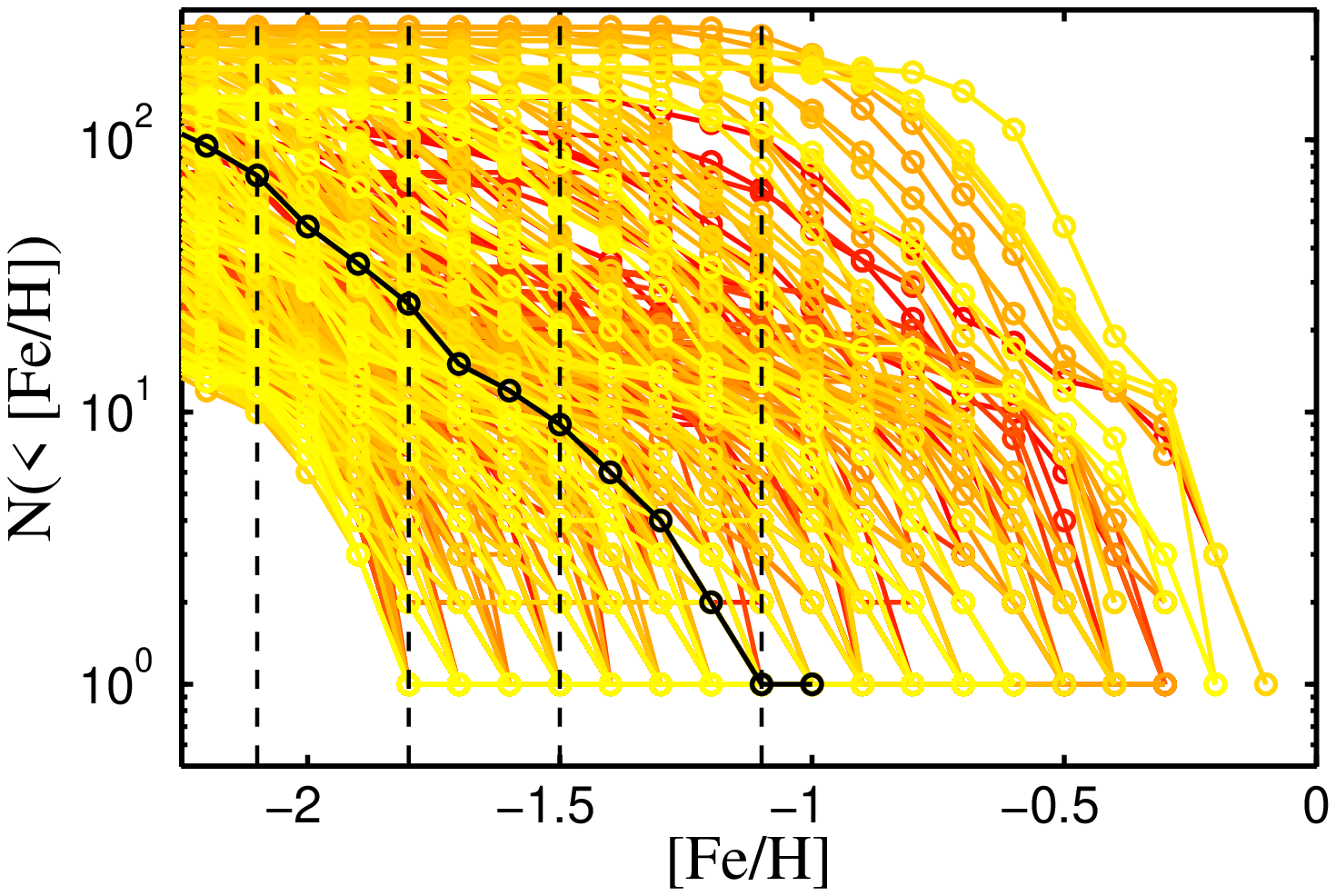}
\caption{The  different colored  lines show  the cumulative  number of
  satellite  galaxies  as a  function  of  absolute V-band  magnitude,
  M$_{\rm     v}$,    (left     panel)    and     mean    metallicity,
  $\langle$[Fe/H]$\rangle$, extracted from a set of 500 models used to
  train the model emulators. The black solid line shows the cumulative
  functions     obtained    from     the    fiducial     model    (see
  \tabref{tab:params}).  The vertical  black dashed lines indicate the
  values chosen to sample the respective cumulative functions.}
\label{fig:hists}
\end{figure*}

In this Section we briefly describe the $N$-body simulations analyzed
in this work, and also provide a brief summary of the main
characteristics of our semi-analytical model ChemTreeN. For a detailed
description of our numerical methods, we refer the reader to
\citet[][]{tumlinson2}, hereafter T10.

\subsection{$N$-body simulations}
\label{sec:sims}

Four  separate simulations  of the  formation of  Milky  Way-like dark
matter  halos are  analyzed in  this work.   The simulations  were run
using  Gadget-2  \citep{springel2005}  on  a local  computer  cluster.
Milky  Way-like   halos  were  first  identified   in  a  cosmological
simulation with  a particle resolution of $128^{3}$  within a periodic
box of side  7.32 $h^{-1}$ Mpc.  The WMAP3  cosmology \citep{wmap} was
adopted,  with matter  density  $\Omega_{m} =  0.238$, baryon  density
$\Omega_{b}= 0.0416$, vacuum energy density $\Omega_{\Lambda}= 0.762$,
power  spectrum  normalization $\sigma_{8}  =  0.761$, power  spectrum
slope $n_{s} = 0.958$, and  Hubble constant $H_{0} = 73.2$ km s$^{-1}$
Mpc$^{-1}$.      The    candidates     were    selected     to    have
gravitationally-bound dark matter halos with virial masses of M$_{200}
\approx 1.5 \times 10^{12}$~M$_{\odot}$  at $z=0$ and no major mergers
since $z  = 1.5$  - 2.   These Milky Way-like  dark matter  halos were
subsequently re-simulated  at a resolution of $512^{3}$  by applying a
multi-mass particle  ``zoom-in'' technique.  At  this resolution, each
dark matter  particle in the  highest-resolution region has a  mass of
$M_{p} = 2.64 \times 10^{5}$ M$_{\odot}$.  Snapshots were generated at
intervals of 20 Myr before $z=4$ and  at 75 Myr intervals from $z = 4$
to $z=0$.  A  six-dimensional friends-of-friends algorithm \citep{fof}
was  applied to  identify dark  matter  halos in  each snapshot.   The
gravitational softening length was 100 comoving pc in all simulations.
The main properties  of the resulting dark matter  halos are listed in
\tabref{tab:sims}.

\begin{deluxetable*}{@{}ccccc}
  \tabletypesize{\footnotesize}                     \tablecaption{Model
    Parameters.\label{tab:params}}                  \tablewidth{360pt}
  \tablehead{\colhead{Parameter}   &    \colhead{Fiducial   Value}   &
    \colhead{Range}  &   \colhead{Description}  &  \colhead{Explored}}
  \startdata
  $z_{\rm r}$ & 10 & 5 -- 19 & Epoch of re-ionization & Yes \\
  $f_{\rm bary}$ & 0.05 & 0 -- 0.2 & Baryonic mass fraction & Yes \\
  $f_{\rm esc}$ & 50 & 0 -- 110 & Escape factor of metals & Yes \\
  $\epsilon_{*}$ & $1$ & 0.2 -- 1.8
  & Star formation efficiency  ($10^{-10}$ yr$^{-1}$) & Yes\\
  $m^{\rm II}_{\rm Fe}$ & 0.07 & 0.04 -- 0.2 &  SN II iron yield ($M_{\odot}$) & Yes\\
  $f_{\rm Ia}$ & 0.015 & 0.005 -- 0.03 & SN Ia probability & Yes \\
  $\epsilon_{\rm SN}$ & 0.0015 & 0.0005 -- 0.006 & SNe energy coupling
  & Yes \\
  $m^{\rm  Ia}_{\rm  Fe}$  &  0.5  &  $\cdots$  &  SN  Ia  iron  yield
  ($M_{\odot}$) & No
\enddata

\end{deluxetable*} 

\subsection{Galactic chemical evolution model and particle tagging}

 In this work we  use the semi-analytical model ChemTreeN, coupled
  to  cosmological simulations, to  follow the  time evolution  of the
  baryonic  component  of  the  stellar  halos.  In  this  context,  a
  semi-analytic  model  consists  of  a set  of  coupled  differential
  equations  describing  the  evolution  of  baryons,  including  star
  formation and  chemical enrichment,  and derives its  mass accretion
  histories  and  spatial   information  from  the  underlying  N-body
  simulations.  Processes  such as  star formation, stellar  winds and
  chemical enrichment are introduced in the model through differential
  equations  that  are  controlled  via  a  set  of  adjustable  input
  parameters.   These parameters  are commonly  set  to simultaneously
  match a range of observable quantities such as the galaxy luminosity
  functions  \citep[e.g.][]{2006MNRAS.370..645B} or  a set  of scaling
  relations \citep[e.g.][]{2011ApJ...727...78K}.  

  The general approach used in our  models is to assume that each dark
  matter  halo found  in  the simulations,  and  followed through  the
  merger  tree,  possesses  gas   that  has  been  accreted  from  the
  intergalactic medium  (IGM), that this  gas forms stars,  that these
  stars return  metals and energy to  the host halo and  to the larger
  environment, and that future generations  of stars form with the now
  metal-enriched  gas.  

  For  every halo  in the  simulation, the  star formation  history is
  calculated using  10 timesteps  between each redshift  snapshot.  At
  each timestep a  star formation ``parcel'' is created  with a single
  initial mass, metallically, and IMF.  The metallicity for the parcel
  is  derived from  the  present gas  metallicity.   Each parcel  thus
  represents   a   single-age  stellar   population   with  a   unique
  metallicity.   When   halos  merge,  their  lists   of  parcels  are
  concatenated.   To  explore the  spatial,  kinematic, and  dynamical
  properties of  stellar populations in the  resulting halos ``stars''
  are assigned  to dark matter  particles in the N-body  simulation at
  each snapshot output.   This is done by selecting  a fraction of the
  most bound particles in each halo.  The star formation that occurred
  between a given  snapshot and the previous one  is identified and an
  equal fraction of the newly formed stars is assigned to each of the
  selected   particles.    In  this   work   only   the  $10\%$   most
  gravitationally  bound particles  in  each halo  are considered,  in
  order  to  approximate the  effect  of  stars  forming deep  in  the
  potential well of the galaxy where dense gas would be most likely to
  collect.

  What follows is a brief description of the physical prescriptions in
  ChemTreeN  that are  most relevant  for this  work. For  more details
  about  this   model,  we  direct   readers  to  our   previous  work
  \citep{tumlinson1, tumlinson2}.

\begin{itemize}
\item {\it Baryon Assignment}

  The number of stars that  a galaxy has formed throughout its history
  strongly depends on the amount of gas it contained.  It is therefore
  important to define a  prescription to model baryonic accretion into
  dark matter halos.  Our models adopt a prescription based on that of
  \citet{bj05} that heuristically takes  into account the influence of
  a photoionizing background from  the aggregate star formation in all
  galaxies.   This model  assigns a  fixed mass  fraction  of baryons,
  $f_{\rm  bary}$,  to all  dark  matter  halos before  re-ionization,
  $z_{\rm r}$.   After $z_{\rm r}$,  gas accretion and  therefore star
  formation  are suppressed  in small  halos with  a  circular velocity
  below $v_{\rm  c} = 30$  km s$^{-1}$.  Between  $v_{\rm c} =  30$ km
  s$^{-1}$  and 50 km  s$^{-1}$, the  assigned baryon  fraction varies
  linearly  from  0 to  $f_{\rm  bary}$.   This  baryon assignment  is
  intended  to capture  the IGM  ``filtering mass''  \citep{gne} below
  which halos are too small to retain baryons that have been heated to
  $T \gtrsim 10^{4}$ K by global re-ionization.

\item {\it Star Formation Efficiency}

  Stars are  formed with  a constant efficiency,  $\epsilon_{*}$, such
  that the  mass formed into  stars $M_{*} = \epsilon_{*}  M_{\rm gas}
  \Delta  t$  in  time   interval  $\Delta  t$.   The  star  formation
  efficiency is  equivalent to a timescale,  $\epsilon_{*} = 1/t_{*}$,
  on which baryons are converted  into stars.  

\item {\it Stellar Initial Mass Function}

An invariant stellar  initial mass function (IMF) at  all times and at
all metallicities  is assumed.  The  invariant IMF adopted is  that of
\citet{kroupa2001}, $dn/dM \propto (m/M_{\odot})^{\alpha}$, with slope
$\alpha =  -2.3$ from  $0.5$ -- $140$  $M_{\odot}$ and slope  $\alpha =
-1.3$ from $0.1$  -- $0.5$ $M_{\odot}$.  

\item {\it Type Ia SNe}

  Type  Ia SNe  are  assumed to  arise  from thermonuclear  explosions
  triggered by  the collapse of a  C/O white dwarf  precursor that has
  slowly accreted  mass from a  binary companion until it  exceeds the
  1.4  $M_{\odot}$ Chandrasekhar  limit.  For  stars that  evolve into
  white  dwarfs as binaries,  the SN  occurs after  a time  delay from
  formation that is roughly equal to the lifetime of the least massive
  companion.   In our  models, stars  with  initial mass  $M =  1.5-8$
  $M_{\odot}$ are  considered eligible to  eventually yield a  Type Ia
  SN.   When stars in  this mass  range are  formed, some  fraction of
  them, $f_{Ia}$, are assigned status as  a Type Ia and given a binary
  companion   with   mass  obtained   from   a  suitable   probability
  distribution  \citep{gregren}. The  chemical  evolution results  are
  sensitive  to the  SN Ia  probability normalization,  $f_{Ia}$. The
  fiducial  value of  this parameter  is fixed  by normalizing  to the
  observed  relative rates  of  Type II  and  Type Ia  SNe for  spiral
  galaxies  in the  local universe  \citep{1994ApJS...92..487T}.  This
  normalization gives a ratio of SN II to Ia of 6 to 1.

\item {\it Chemical Yields}

ChemTreeN tracks the time evolution of galaxies' bulk metallicities by
considering Fe  as the proxy reference  element. For Type  Ia SNe with
1.5 -- 8 $M_{\odot}$ the models adopt the W7 yields of \citet{nomo1997}
for Fe, with 0.5 $M_{\odot}$ of Fe  from each Type Ia SN.  Type II SNe
are assumed  to arise from  stars of 10  to 40 $M_{\odot}$,  with mass
yields  provided by  \citet{2009ApJ...690..526T}.  They  represent the
bulk yields  of core-collapse SNe  with uniform explosion energy  $E =
10^{51}$ ergs.   These models have $M  = 0.07$ --  $0.15$ $M_{\odot}$ Fe
per event.

\item{\it Chemical and Kinematic Feedback}

One possible cause of  the observed luminosity-metallicity ($L$-$Z$)
relation for  Local Group dwarf  galaxies is SN-driven mass  loss from
small dark matter halos \citep{2003MNRAS.344.1131D}.   To model  this physical
mechanism, ChemTreeN tracks mass loss  due to SN-driven winds in terms
of the number of SNe per timestep in a way that takes into account the
intrinsic time variability in the  star formation rate and rate of SNe
from a stochastically sampled IMF.  At each timestep, a mass of gas
\begin{equation}
M_{\rm lost} = \epsilon_{\rm SN} \sum_{i} \dfrac{N^{i}_{\rm SN}
E^{i}_{\rm SN}} {2v_{\rm circ}^{2}}
\end{equation}

becomes  unbound and is  removed permanently  from the  gas reservoir.
Here  $v_{\rm circ}$  is the  maximum circular  velocity of  the halo,
$N_{\rm SN}$  is the number of  SNe occurring in a  given timestep and
$E_{\rm  SN}$ is  the energy  released by  those SNe.   The  only free
parameter,  $\epsilon_{\rm  SN}$, expresses  the  fraction  of the  SN
energy that is converted to kinetic  energy retained by the wind as it
escapes.  The sum  over index i sums over all  massive stars formed in
past timesteps  that are just  undergoing an explosion in  the current
timestep.  Note that this approach allows for variations in the number
and energy  of SNe  from timestep to  timestep. The selective  loss of
metals that  should arise when SNe  drive their own ejecta  out of the
host  galaxy  is  captured  by  the  parameter  $f_{\rm  esc}$,  which
expresses the increased metallicity  of the ejected winds with respect
to the ambient interstellar medium.  At each timestep, a total mass in
iron $M^{\rm Fe}_{\rm lost}$ is  removed from the gas reservoir of the
halo:
\begin{equation}
  M^{\rm Fe}_{\rm lost} = f_{\rm  esc} M_{\rm lost} \dfrac{M^{\rm Fe}_{\rm ISM}}{M_{\rm gas}}
\end{equation}

where $M^{\rm Fe}_{\rm  ISM}$ is the total mass of  iron in the ambient
interstellar  medium, $M_{\rm  gas}  \times 10^{\rm  [Fe/H]}$ .   This
prescription ensures  that, on average, the ejected  winds are $f_{\rm
  esc}$  times  more  metal-enriched  than  the  ambient  interstellar
medium.  Alternatively, the fraction of  metal mass lost from the halo
is  $f_{\rm esc}$ times  higher than  the total  fraction of  gas mass
lost.

\item {\it Isochrones and Synthetic Stellar Populations}

  To compare these model halos to observational data on the real Milky
  Way  and its  dwarf satellites,  it  is necessary  to calculate  the
  luminosities   and  colors  of   model  stellar   populations  using
  pre-calculated  isochrones and  population  synthesis models.   Each
  star  formation parcel  possesses a  metallicity, age,  and  a total
  initial mass distributed according  to the assumed IMF.  These three
  quantities  together uniquely  specify an  isochrone and  how  it is
  populated.  The models adopt  the isochrones  of \citet{girardi2002,
    girardi2004}  for  the  UBVRIJHK  and SDSS  {\it  ugriz}  systems,
  respectively,     as    published     on     the    Padova     group
  website\footnote{\url{http://stev.oapd.inaf.it/}}.     The    lowest
  available metallicity  in these isochrones  is [Fe/H] =  -2.3. Thus,
  this  value is  used  to represent  stellar  populations with  lower
  metallicities.

\end{itemize}

\tabref{tab:params} summarizes the  numerical values of the parameters
used for  our fiducial  models, as  well as the  range of  values over
which they are allowed to vary.

\section{Statistical Methods}
\label{sec:mod_emu}

 In this  Section we  describe the  statistical methods  that are
  applied  throughout the  text. We  start by  reviewing  the Gaussian
  process  model  emulation  technique  introduced in  G12.   We  then
  describe  in  Section~\ref{sec:sa_an}   a  novel  application  of  a
  technique  known as  Sensitivity  Analysis which  will  allow us  to
  characterize  the input-output  relationship of  our chemo-dynamical
  model ChemTreeN. 

\subsection{Gaussian Process model emulator}
\label{sec:gp}

In what  follows we briefly describe  how to train  a Gaussian process
model   emulator  \citep{OHag:2006,   Oakl:Ohag:2002,  Oakl:Ohag:2004,
  Kenn:OHag:2000}  and  we  refer  the  reader to  \citet{G12}  for  a
detailed      description     of     the      procedure     \citep[see
also][]{2010MNRAS.407.2017B}.    An   emulator   is   constructed   by
conditioning a Gaussian process prior on a finite set of model outputs
(or mock  observables), collected  at points dispersed  throughout the
parameter space.   Once the  emulator is trained  it can  rapidly give
predictions  of the  model outputs,  and an  attendant measure  of its
uncertainty, at any  point in the parameter space.   In other words it
acts as a statistical model of our much more computationally-expensive
ChemTreeN  model.   Numerical   implementations  of  Gaussian  process
emulators are computationally efficient, making it feasible to predict
vast numbers of model outputs in a short period of time.

  A  Gaussian  process  is  a stochastic  process,  all  of  whose
  finite-dimensional marginal distributions are multivariate normal --
  i.e., a  single sample  is normally distributed,  a pair  of samples
  have a two dimensional  joint multivariate normal distribution, etc.
  Let $\mathcal{D} = \{\vx_1, \ldots,  \vx_n\}$ be a set of $n$ points
  in  a  $p$-dimensional  input  parameter  space. We  will  refer  to
  $\mathcal{D}$  as  design.   In  this  work  a  typical  element  of
  $\mathcal{D}$  is  a 7-dimensional  input  parameter  vector $\vx  =
  (z_{\rm  r},~  f_{\rm bary},~  f_{\rm  esc},~ \epsilon_{*},~  m^{\rm
    II}_{\rm  Fe},~ f_{\rm Ia},~  \epsilon_{\rm SN},~  m^{\rm Ia}_{\rm
    Fe})$.    Let  $\mathbf{Y}   =  \{y_1,   \ldots,  y_n\}$   be  the
  corresponding  set of  $n$  training values  representing the  model
  output at the  design locations.  For example, a  typical element of
  $\mathbf{Y}$  is  the  cumulative  number  of  Milky  Way  satellite
  galaxies at $M_{\rm  v} \leq -5$, modeled by  ChemTreeN at $\vx \in
  \mathcal{D}$. The posterior distribution defining our emulator is
\[ 
\mathcal{P}(\mathbf{Y}\mid\vx,\theta)  \sim \mbox{GP}\left(m(\vx,
  \mathbf{\theta}), \Sigma(\vx, \mathbf{\theta})\right),
\]
with
\begin{align}
  \label{eqn-emu-mean-var}
  m(\vx) &= \mathbf{h}(\vx)^{T}\hat{\beta} +
  \mathbf{k}^{T}(\vx) \mathbf{C}^{-1} ( \mathbf{Y} - {\bf H} \hat{\beta}), \notag \\
  {\Sigma}(\vx_i, \vx_j) &= c(\vx_i, \vx_j)  - \mathbf{k}^{T}(\vx_i) \mathbf{C}^{-1} \mathbf{k}(\vx_j) + \Gamma(x_i, x_j), \notag\\
  \mathbf{C}_{ij} &= c(\vx_i, \vx_j) \\
  \Gamma(x_i,   x_j)    &=   \left(   \mathbf{h}(\mathbf{x_i})^{T}   -
    \mathbf{k}^{T}(\mathbf{x_i})\mathbf{C}^{-1} {\bf H}\right)^{T}
  \left({\bf H}^{T} \mathbf{C}^{-1} {\bf H}\right)^{-1} \notag \\
  &\left(\mathbf{h}(\mathbf{x_j})^{T} -
    \mathbf{k}^{T}(\mathbf{x_j})\mathbf{C}^{-1}{\bf H} \right), \notag \\
  \mathbf{k}(\vx)^{T} &= \left( c(\vx_1,  \vx) , \ldots, c(\vx_n, \vx)
  \right), \notag.
\end{align}

Here, $m(\vx)$ is the posterior mean at $\vx$,
$\Sigma(\vx_i, \vx_j)$ is the posterior covariance
between points $\vx_i$ and $\vx_j$, $\mathbf{C}$ is
the $n \times n$ covariance matrix of the design $\mathcal{D}$,
$\hat{\beta}$ are the maximum-likelihood estimated regression
coefficients, $\mathbf{h}$ the basis of regression functions and
${\bf H}$ the matrix of these functions evaluated at the training
points. The elements of the vector $\mathbf{k}(\vx)$ are the
covariance of an output at $\vx$ and each element of the training set.

To construct an emulator we need to fully specify our Gaussian process
 by choosing forms for the prior mean and covariance functions. We
model the prior mean by linear regression with some basis of functions
$\mathbf{h}(\vx)$. We use $\mathbf{h}(\vx)  = \{ 1 \}$ for simplicity.
We specify a power exponential form for the covariance function,
\begin{equation}
   \label{eqn-emu-cov}
   c(\mathbf{x}_i, \mathbf{x}_j) = \theta_0 \exp\left(-\frac{1}{2}
     \sum_{k=1}^{p} \left(\frac{|x_i^{k} -
       x_j^{k}|}{\theta^{k}}\right)^{\alpha}\right) + \delta_{ij} \theta_{N}. 
\end{equation}
Here $\theta_0$ is the marginal variance, the $\theta^{k}$ set
characteristic length scales in each dimension in the parameter space
and $\theta_N$ is a small term, usually called a nugget, added to
ensure numerical convergence or to model some measurement error in the
code output.  The exponent $1 \le \alpha < 2$ sets the roughness of
the functions generated by the stochastic process.  For this analysis
we pick a value just less than $2$ which ensures smoothness.  The shape of the covariance
function sets how correlations between pairs of outputs vary as a
function of the distances between the corresponding input vectors in
the parameter space.  The scales in the covariance function
$\theta^{k}$ are estimated from the training data using maximum
likelihood methods \citep{Rasmussen05}.

A  maximin Latin  Hyper  Cube (LHC)  design  is used  to generate  the
training  locations in  the  parameter space.   This  is an  efficient
design  for  space-filling   in  high  dimensional  parameter  spaces.
\citep{Sack:Welc:Mitc:Wynn:1989,  Sant:Will:Notz:2003}.  LHC sampling
scatters $N$ points  in a $p$-dimensional cube in such  a way that all
one   and    two-dimensional   marginals   have    $N$   approximately
uniformly-spaced  points,  while  a   regular  grid  would  have  only
$N^{1/p}$ or $N^{2/p}$ distinct marginal points, respectively.

Following \citet{G12} (see also B10), to compare the emulated model
output to experimental data we define a univariate implausibility
measure,
\begin{equation}
  \label{eqn-implaus-scalar}
  I^{2}(\vx)  = \frac{({m}(\vx) - E[Y_f])^2}  {{\Sigma}(\vx, \vx) + V[Y_f]},
\end{equation}
where  $Y_f$   represents  the   experimental  or  field   data  (real
observables) that we  seek to compare our model  against, $E[Y_f]$ the
expected value of $Y_f$  and $V[Y_f]$ the observational uncertainties.
Large  values   of  $I(\vx_{t})$  indicate   that  the  input
parameter vector $\vx_{t}$ is unlikely to give a good fit to
the  observable  data.   Note  that $I(\vx)$  is  a  unit-less
quantity. 

 The implausibility  can  be easily  generalized  to account  for
  multivariate  outputs.  Consider a  $t$-dimensional vector  of model
  outputs  $\mathbf{y(\vx)}  =  \{y_1,  \ldots,  y_t  \}$.  Here,  the
  elements  of  $\mathbf{y(\vx)}$  are,  for example,  the  cumulative
  number of  Milky Way satellite  galaxies at $t$ different  values of
  $M_{\rm v}$,  modeled by ChemTreeN  at $\vx \in  \mathcal{D}$.  We
extend our training set to be the $t \times n$ matrix $\mathbf{Y} = \{
\mathbf{y(\vx_1)}, \ldots, \mathbf{y(\vx_n)} \}$.  We define the joint
implausibility   $J(\vx)$    for   observables   $\mathbf{Y_f}$   with
measurement    variance     $V[\mathbf{Y_f}]$    and    mean    values
$E[\mathbf{Y_f}]$:
\begin{multline}
  \label{eqn-implaus-joint}
  J^{2}(\vx) =
    \left(E[\mathbf{Y_f}]- \mathbf{{m}}(\vx) \right)^{T} \\
    \left(\mathbf{K}(\vx) + I \cdot V[\mathbf{Y_f}] \right)^{-1}
    \left(E[\mathbf{Y_f}]- \mathbf{{m}}(\vx) \right),
\end{multline}
where $\mathbf{K}(\mathbf{x})$ represents  the emulated $t \times t$
dimensional covariance  matrix between the model outputs  at the point
$\vx$   in  the   design  space   and  $\mathbf{{m}}(\vx)$   is  the
$t$-dimensional   emulator  mean  vector.    This  covariance-weighted
combination of the multiple  observables gives a reasonable indication
of which input values $\vx$  are predicted by the emulator to
lead to model predictions close to the observed values $\mathbf{Y}$.
Note that $J(\vx)$ is a  $p$-dimensional scalar function, with $p$ the
number  of considered  input  parameters.  For  the optimal  parameter
vector $\vx$,  the quantity $J(\vx)^2$ has  approximately a $\chi^2_t$
distribution  (see discussion  in G12),  leading in  the usual  way to
confidence sets in the input space.  Following G12, we consider $75\%$
confidence sets.

\begin{figure*}
\centering
\includegraphics[width=180mm,clip]{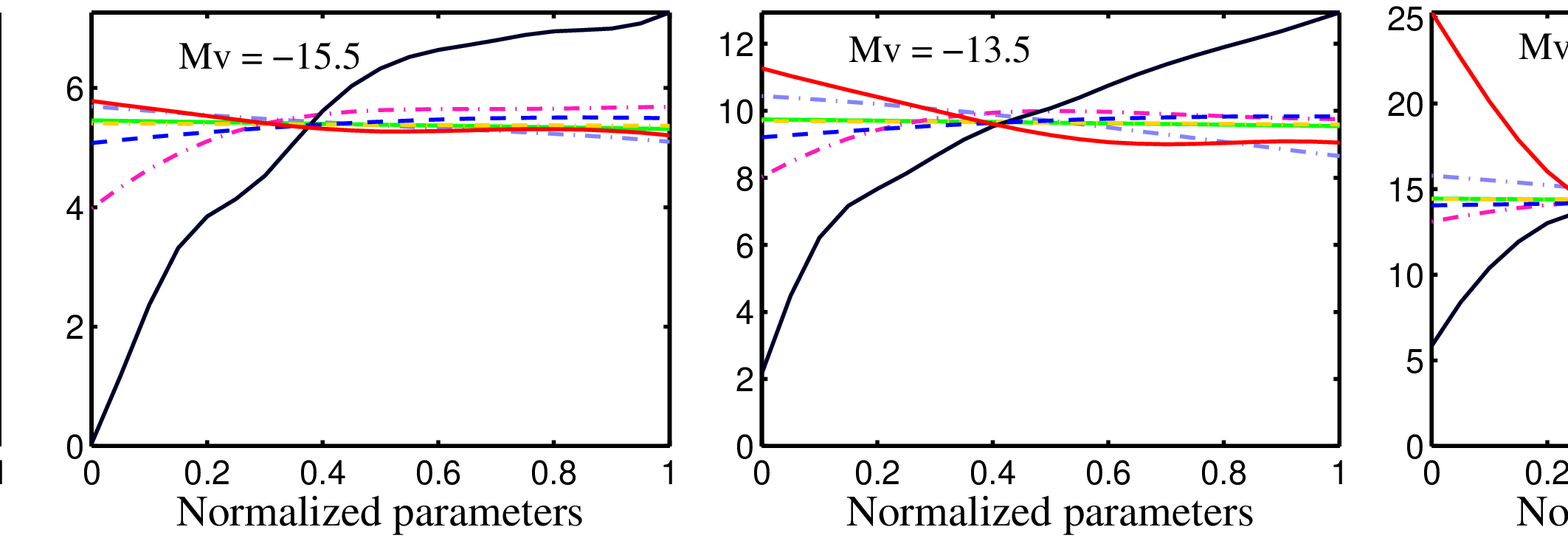}
\\
\includegraphics[width=180mm,clip]{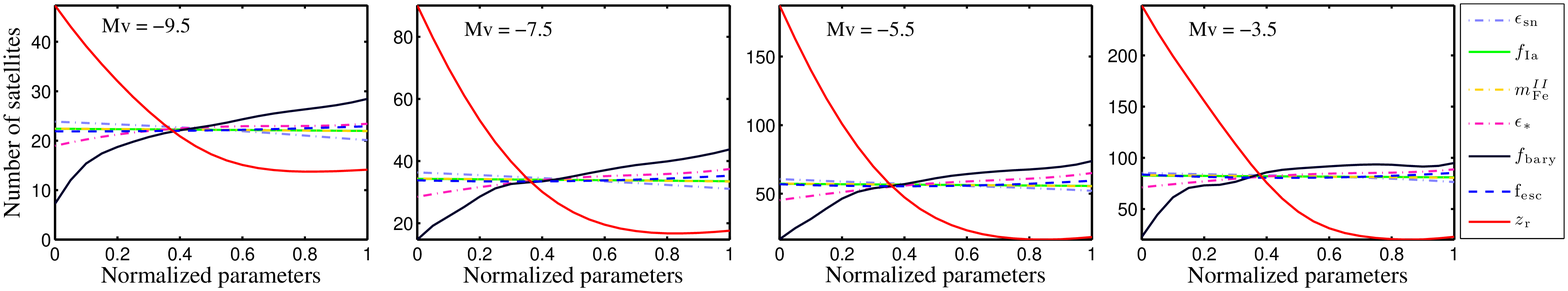}
\caption{Main  effects  obtained  from  a  seven-dimensional  Gaussian
  process   model  emulator   of  ChemTreeN,   where   each  dimension
  corresponds  to  a  different  input  variable.   The  results  were
  obtained using the simulation labeled MW1. The different panels show
  the results for different mock  observables. From left to right, the
  columns  correspond  to  eight  different  bins  of  the  luminosity
  function.  The corresponding mock observable is indicated on the top
  left corner of  each panel.  On each panel, the  lines show the main
  effect associated  with a different input variable,  as indicated in
  the legend  located at  the bottom right  corner. The range  of each
  input  variable  has  been  normalized to  the  corresponding  total
  extent,  indicated in  \tabref{tab:params}. From  this figure  it is
  possible to  infer what parameters are most  important to explaining
  the variability observed on each mock observable.  Note as well that
  some parameters, such as $m_{\rm  Fe}^{\rm II}$ and $f_{\rm Ia}$, do
  not  show a  strong influence  on the  values of  the  selected mock
  observables.}
\label{fig:maineff}
\end{figure*}

\subsection{Sensitivity Analysis}
\label{sec:sa_an}

Semi-analytical models are  conceptually very simple. Individually, it
may  seem straightforward  to  forecast how  variations  of the  input
parameters associated with an  adopted prescription can affect a given
mock observable.    However,  the   complex  nonlinear   couplings  between
different physical  processes, in addition to  the high dimensionality
of the problem, can make  this into an extremely challenging task.  It
is  therefore desirable  to  implement techniques  that  allow one  to
statistically   characterize  the   relationship  between   the  input
parameters  and each  mock observable.  A  sensitivity analysis  is  a very
powerful  technique  for   dissecting  computer  models  \citep{sa_1},
providing information  about which  parameters are the  most important
and most likely  to affect the prediction of  any given observable. It
can  also  allow  us  to  simplify  our  model  by  identifying  input
parameters  that have  little or  no effect  on the  available  set of
outputs or mock observables. To carry  out a sensitivity analysis  (SA) on
ChemTreeN, we follow  the approach described by \cite{X}.  Below we provide a short
description of  the method and we  refer the reader to  their work for
more details.

The main goal of this analysis is to decompose the input-output
relationship of ChemTreeN into a set of orthogonal quantities called
main effects and interactions.  These characterize how an output
responds to variations of only a subset of input variables, allowing
us to obtain a decomposition of the total variance observed. Main
effects are those quantities in this expansion associated with
variations of single input variables, and interactions or joint
effects are those quantities associated with variations of two or more
input variables.

\begin{figure}
\centering
\includegraphics[width=89mm,clip]{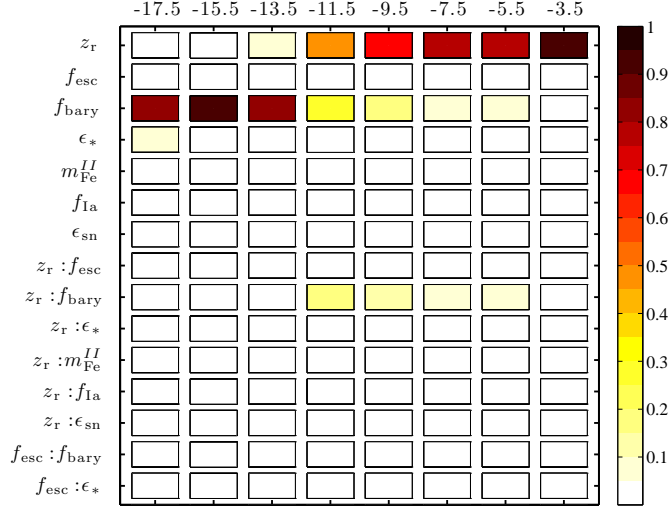} 
\caption{ANOVA  decomposition (see  \ref{sec:sa_an})  obtained from  a
  seven  dimensional  Gaussian process  model  emulator of  ChemTreeN,
  where each dimension corresponds to a different input variable.  The
  results  were  obtained  using  the  simulation  labeled  MW1.   The
  different columns correspond  to different mock observables, whereas rows
  are associated with either main  effects or interactions. From left to
  right, the  columns correspond to  different bins of  the luminosity
  function.   We  only   consider  up   to   two-variable  interaction
  effects. Note that, for  simplicity, not all interaction effects are
  shown.  The  different colors indicate  the percentage of  the total
  variance  that can be  explained by  the corresponding  effect.  The
  total  variance  associated with  each mock observable  (column) has  been
  normalized  to  one.  This  graphical  representation  of the  ANOVA
  decomposition allows us to quickly identify what input parameters are
  more  important  in  explaining  the variability  observed  on  each
  observable.}
\label{fig:fanova}
\end{figure}

To  apply a  sensitivity analysis  it is  necessary to  densely sample
ChemTreeN  over the  whole range  of interest  of its  input parameter
space.   With   the  ChemTreeN   code,  doing  this   rapidly  becomes
computationally  prohibitive  as   the  dimensionality  of  the  input
parameter space increases. Thus, we will perform a sensitivity analysis on the posterior
mean associated  with the  corresponding Gaussian process model  emulator, $\vm(\vx,
\mathbf{\theta})$. In what follows,  for simplicity we will refer to
the  conditional mean  as $m(\vx)$  and  assume $t=1$,  i.e., we  will
consider  a  single mock  observable  (see Section~\ref{sec:gp}).   Let  us
consider the  effect of the  subset of input variables  $\vx_e$, where
$e$ denotes the  indexes of the variables we  are interested in.  Note
that  $\dim(\{\vx_{e},\vx_{-e}\})  =  p$.   The simplest  approach  to
estimate the  effect associated with  $\vx_e$ is to fix  the remaining
variables  $\vx_{-e}$  at  a  given  value,  e.g.,  their  mid-ranges.
However,  the effect  associated with  the variables  in  $\vx_{e}$ is
likely  to  depend on  the  values  chosen  for $\vx_{-e}$.   Instead,
effects  are  defined   by  averaging  $m(\{\vx_{e},\vx_{-e}\})$  over
$\vx_{-e}$, ${\otimes_{\chi_j:~j\ne e}}$

\begin{equation}
\label{eq:effects}
\overline{m}_{\vx_{e}}(\vx_{e}) = \int\limits_{\otimes_{\chi_j:~j~\ne~e}} m(\vx_{e},\vx_{-e}) \, \prod_{j~\ne~e} \omega_{j}(x_j) \, \mathrm{d}x_{j},
\end{equation}
where  $  \omega_{j}(x_j)$,  with $j  =1,\ldots, ~p$,  are a  set  of
orthogonal weight functions, often chosen to be a uniform
distributions, and
\begin{equation}
\chi = \otimes^{p}_{j=1}\, \chi_{j}
\end{equation}
represent the domain of $m(\vx)$.

The effects defined in Equation~\ref{eq:effects} can be used to
generate a decomposition of $m(\vx)$ into adjusted effects
involving different numbers of input variables as follows:
\begin{equation}
\begin{split}
m(\vx) = \mu_{0} + \sum_{j=1}^{p} \mu_{j}(x_{j}) +
\sum_{j=1}^{p-1} \sum_{j'=j+1}^{p} \mu_{jj'}(x_{j},x_{j'})+ 
\ldots+\\ \mu_{1,\ldots,p}(x_{1},\ldots,x_{p}),
\end{split}
\end{equation}
where 
\begin{equation}
\mu_{0} = \int\limits_{\chi} m(\vx)~\omega(\vx)~\mathrm{d}\vx
\end{equation}
is an overall average,
\begin{equation}
\mu_{j}(x_{j}) = \overline{m}_{j}(x_{j}) -
\mu_{0},~~{\rm for}~ x_{j} \in \chi_{j}
\end{equation}
is the adjusted main effect of $x_{j}$,
\begin{equation}
\begin{split}
\mu_{jj'}(x_{j},x_{j'}) = \overline{m}_{jj'}(x_{j},x_{j'}) -
\mu_{j}(x_{j}) - \mu_{j'}(x_{j'}) - \mu_{0},\\{\rm for}~ x_{j},x_{j'}
\in \chi_{j} \otimes \chi_{j'},
\end{split}
\end{equation}
is the adjusted joint effect of $x_{j}$ and $x_{j'}$ (often referred to as the first interaction), and so
on. Note that each adjusted effect is just the corresponding effect corrected to remove all lower-order terms. An
important property of the effects is that they are orthogonal with respect to the weight function,
$\omega(\vx)$. This allows one to define a decomposition of the total variance of $m(\vx)$ as follows,
\begin{equation}
\begin{split}
\label{eq:anova}
\int\limits_{\chi} [m(\vx) - \mu_{0}]^{2} ~\omega(\vx)
~{\rm d}\vx = \sum_{j=1}^{p} \int\limits_{\chi_{j}} \mu_{j}^{2}(x_{j})
~\omega_{j} ~{\rm d}x_{j}  + \\ 
\sum_{j=1}^{p-1} \sum_{j'=j+1}^{p} ~ \int\limits_{\chi_{j} \otimes
  \chi_{j'}}
\mu_{jj'}^{2}(x_{j},x_{j'})~\omega_{j}(x_{j})~\omega_{j'}(x_{j'})~{\rm
  d}x_{j}~{\rm d}x_{j'} + \ldots + \\
\int\limits_{\chi} \mu_{1,\ldots,p}^{2}(x_{1},\ldots,x_{p})
\prod_{j=1}^p \omega_{j}(x_{j})~{\rm d}x_{j}.
\end{split}
\end{equation}

This  ANalysis  Of VAriance  (ANOVA)  decomposition  offers  a way  to
quantify the fraction of the total variance, shown on the left side of
Eq.~\ref{eq:anova}, that can be  explained by variations of any single
input variable or by  a combination of two or more. Note that the larger
the  percentage, the  more  sensitive  a mock  observable  is to  the
corresponding input variables.

\begin{figure*}
\centering
\includegraphics[width=180mm,clip]{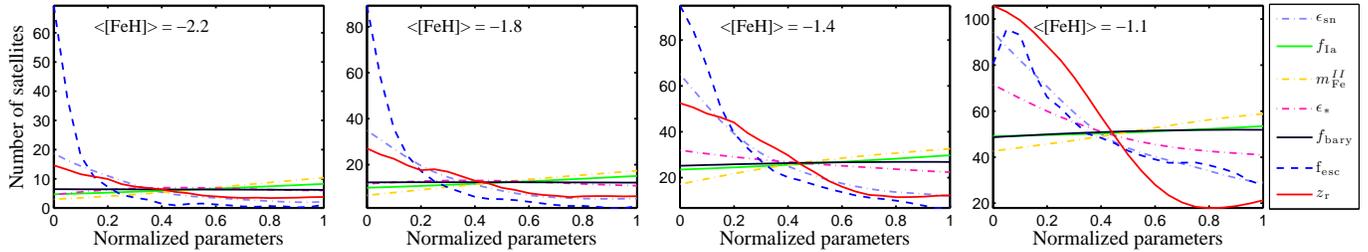}
\caption{As in Figure~\ref{fig:maineff},  now for mock observables obtained
  from the  cumulative number of  satellite galaxies as a  function of
  mean metallicity, $\langle$[Fe/H]$\rangle$.}
\label{fig:maineff_feh}
\end{figure*}

\begin{figure}
\centering
\includegraphics[width=75mm,clip]{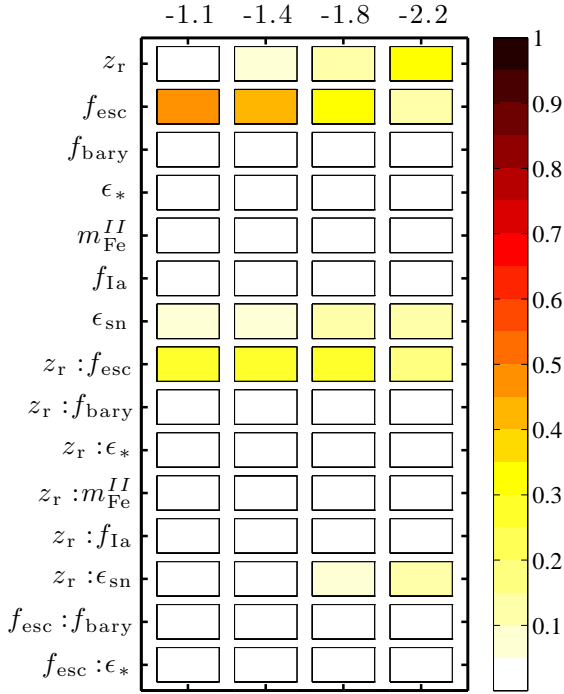}
\caption{As in figure \ref{fig:fanova}, for 4 bins of the cumulative
  number of satellite galaxies as a function of mean metallicity, $\langle$[Fe/H]$\rangle$.}
\label{fig:fanova_feh}
\end{figure}

\section{Dissecting ChemTreeN -- characterizing its input-output relationship}
\label{sec:in-out}

In what follows we will use the statistical tools described in the
previous section to characterize in a quantitative way the
relationships between input parameters and mock observable
quantities produced by the ChemTreeN model.  This is a useful exercise
for several reasons.  First, models of this sort are inexpensive
compared to full-physics cosmological simulations, but take long
enough to run (typically several hours) that sweeping through an
entire range of parameter space is impractical, particularly if said
parameter space has high dimensionality.  As a result, models that
produce a statistically good fit may only represent local maxima in
probability, and thus other comparably good (or better) model
parameter sets, and thus
potentially interesting results, may be missed.  Second, quantifying
the relationships between input parameters (and combinations of
parameters) and output values helps to highlight the most sensitive
relationships between inputs and outputs, and to suggest areas where
further experimentation with model prescriptions (possibly influenced
by more physics-rich numerical simulations) would be particularly
beneficial.  Alternately, this allows us to find input parameters that
have virtually no effect on the output values of interest, and which
can be ignored in future experimentation.  Third, performing such an
analysis for the same ChemTreeN model using different N-body
simulations helps both to identify universal commonalities and to find
outputs where ``implicit'' parameters in the model (e.g., $z=0$ halo
mass or merger history) are important.

In  this  section, we  consider  as  mock  observables the  cumulative
functions of  the surviving satellites as  a function of  {\it a)} the
absolute magnitude  in the  V-band and {\it  b)} the  satellite's mean
metallicity, $\langle  [$Fe/H$]\rangle$. We will refer to  them as the
Luminosity   Function  (LF)   and  the   Metallicity   Function  (MF),
respectively.   The  advantage of  using  values  of these  cumulative
functions as mock observables is that they are easy to emulate and, as
we will show  in what follows, they are  most significantly influenced
by a different set of input parameters.  Once the relationship between
these model outputs and input  parameters has been established, we will
turn  our  attention to  a comparable  set of  real  observables  that have  been
extracted from  a range of  measurements of the Milky  Way's satellite
galaxies.

\begin{figure*}
\centering
\includegraphics[width=87.5mm,clip]{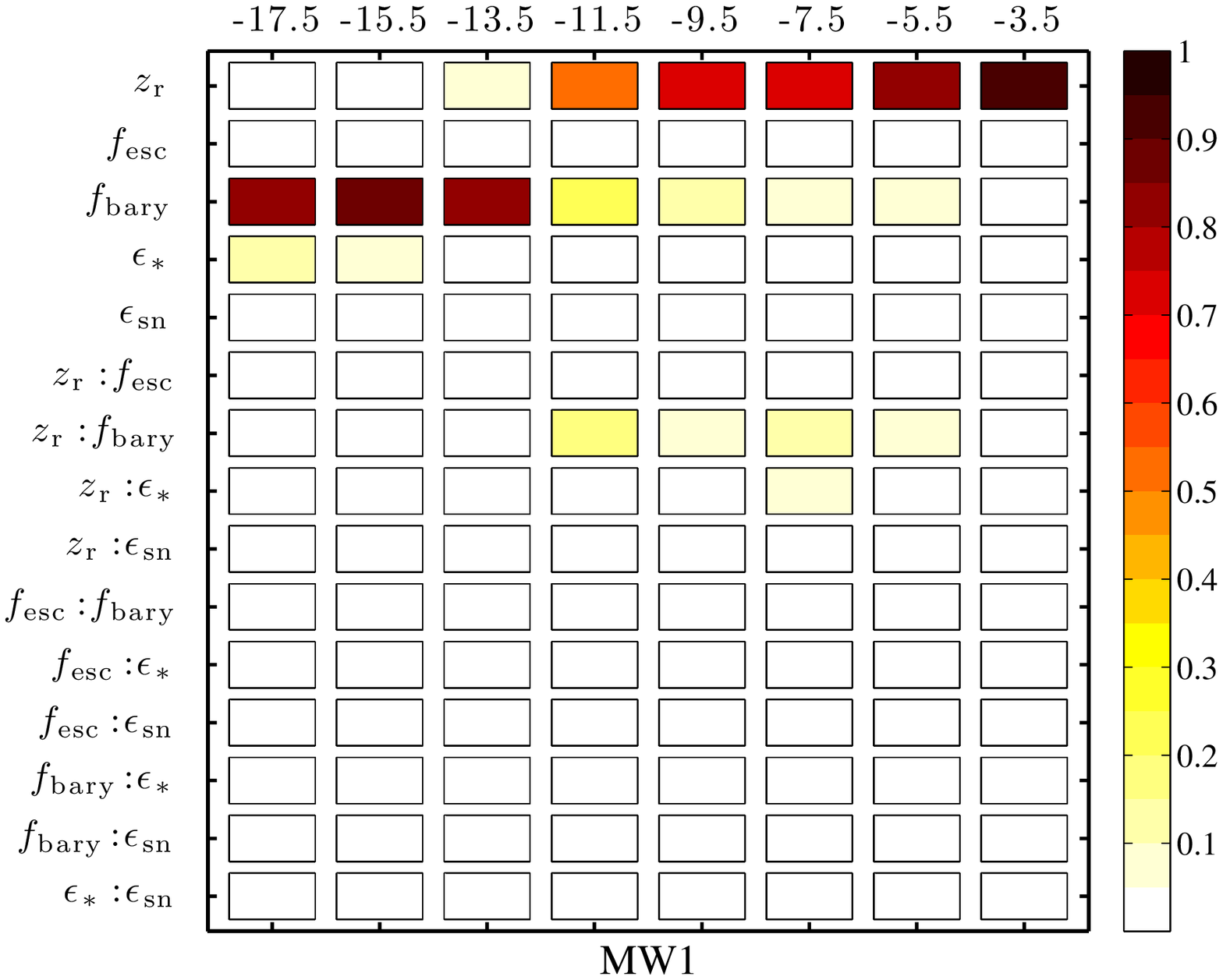}
\includegraphics[width=88.1mm,clip]{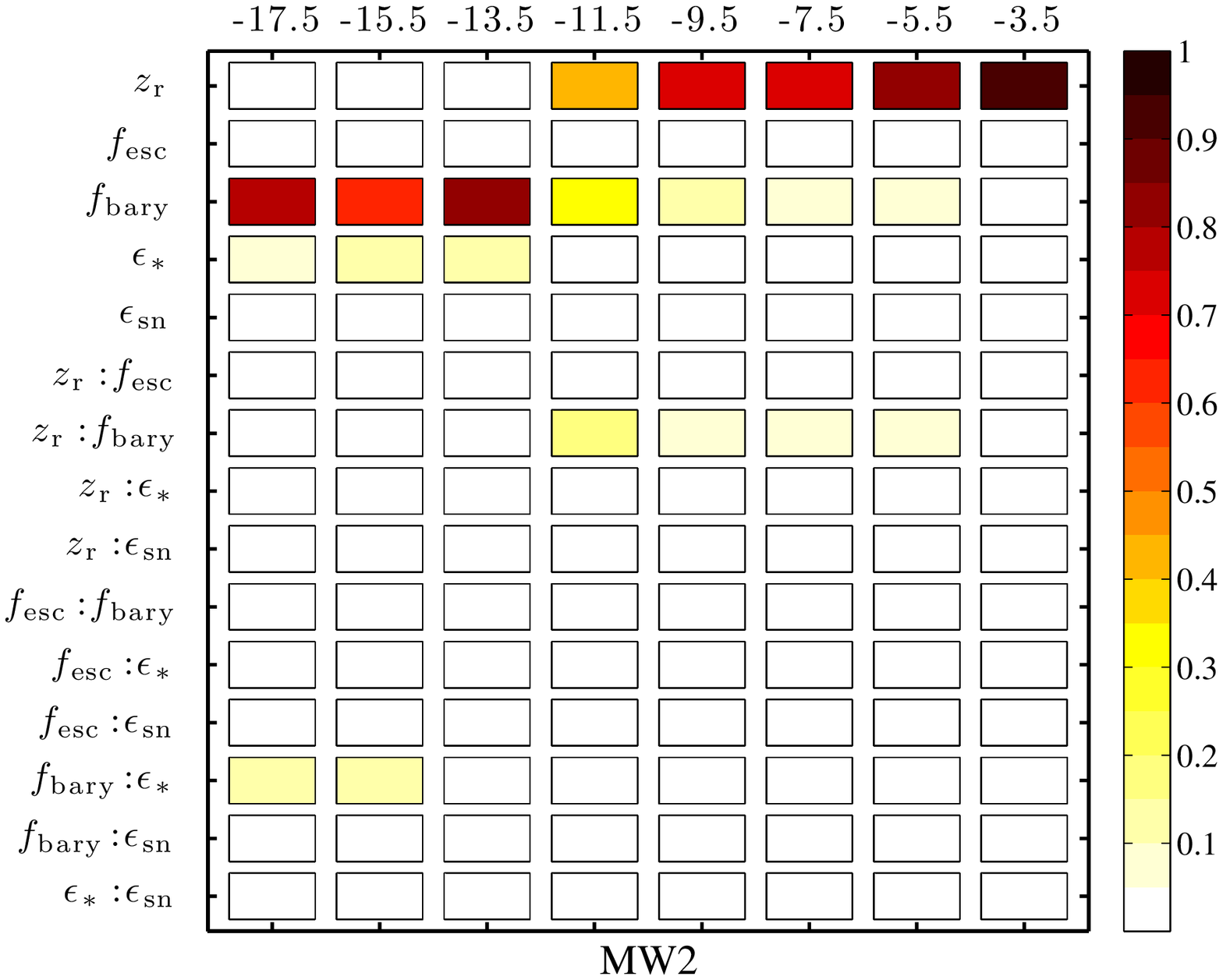}
\caption{ANOVA decomposition obtained after coupling ChemTreeN with
the dark matter-only N-body simulations MW1 (left) and MW2 (right). The same
five dimensional design was used in both cases to create the training
set.  The different columns correspond to different mock observables,
whereas rows are associated with either main effects or interactions.
From left to right, the columns correspond to different bins of the
luminosity function.  Note that the ANOVA decomposition allows us to
characterize the relationship between the input parameters and the
desired model outputs, independently of the corresponding real
observable values and the underlying formation history of our galactic
model.}
\label{fig:anova_diff_dm}
\end{figure*}

The  first step  in  our analysis  consists  of constructing  Gaussian
process model emulators for the  desired set of mock observables.  The
mock observables  selected for emulation are values  of the luminosity
function  and metallicity  function  at different  locations of  their
respective domains.  More precisely,  we emulate the cumulative number
of  satellite galaxies  above a  range of  values of  M$_{\rm  v}$ and
$\langle$[Fe/H]$\rangle$.   The  respective  values are  indicated  in
Figure~\ref{fig:hists} with vertical  black dashed lines. To train
  the  emulators  we created  a  training  set  consisting of  $n=500$
  models. These models were  obtained after running ChemTreeN over 500
  points  dispersed throughout  the  input parameter  space within  the
  ranges specified in \tabref{tab:params}.   This number of points was
  set to adequately balance the  coverage of the input parameter space
  and the associated run time,  and our results are insensitive to the
  number of training points (as long as a sufficient number are used).
  Each model of the training is computed after coupling ChemTreeN with
  the  dark matter-only simulation  MW1.  We  start by  considering a
seven-dimensional  space   of  input  parameters   that  includes  the
parameters flagged  as ``Yes'' in  \tabref{tab:params}.  The resulting
cumulative  functions  of  all  the  training models  are  shown  with
different   colors  in   Figure~\ref{fig:hists}.    As  discussed   in
Section~\ref{sec:gp},   a  Gaussian  process   model  emulator   is  a
statistical model of ChemTreeN that  allows us to obtain predictions of
the desired  model outputs,  and an attendant  measure of their
uncertainty,  at any  point of  the input  parameter space.   Once the
model emulator is  trained it is possible to  compute the main effects
and  first  interactions  (joint   influence  of  two  parameters)  as
described in Section~\ref{sec:sa_an}.

We train two different sets of model emulators.  This is done by using
outputs extracted  either solely from the luminosity  function or from
the  metallicity function.   In Figure~\ref{fig:maineff}  we  show the
main  effects  computed  for   mock  observables  extracted  from  the
luminosity function.  Each panel corresponds to a different bin in the
luminosity function,  from most luminous (top left)  to least luminous
(bottom  right).   Each  line  is  associated with  a  separate  input
parameter, as shown in the key.   Each line tells us how the number of
satellite galaxies at a given value of M$_{\rm v}$ varies as we vary a
single input parameter of ChemTreeN, averaging the model emulator over
the remaining six-dimensional input parameter space.  For the purposes
of comparison,  the range  within which each  parameter is  allowed to
vary has been  normalized from $0-1$.  From the top  left panel we can
observe that,  as expected,  the number of  satellite galaxies  in the
bright  end of  the luminosity  function mainly  depends on  the value
assigned to the baryon  fraction, $f_{\rm bary}$.  For this particular
cosmological  simulation,  the  cumulative  number  of  satellites  at
M$_{\rm v} = -17.5$ could take any value between 0 and $\sim 6$ simply
by varying the  value of this parameter within  the range permitted by
the emulator.  We can also immediately see a much weaker dependence on
the  star   formation  efficiency,  $\epsilon_{*}$.    Note  that  the
remaining  parameters have almost  no effect  on this  particular mock
observable.   As we  move  toward  the faint  end  of the  luminosity
function (i.e.,  less negative values  of M$_{\rm v}$),  the parameter
$f_{\rm bary}$ becomes less important and the redshift of the epoch of
reionization,  $z_{\rm r}$,  starts to  take over.   At M$_{\rm  v} =
-3.5$, in the regime of the  ultra faint dwarf galaxies, the number of
satellite galaxies is  strongly dominated by $z_{\rm r}$,  with a much
weaker dependence on $f_{\rm bary}$.  Again, we find that variation of
the remaining parameters does  not significantly affect the cumulative
number of galaxies in this magnitude bin.

Figure  3 shows  the  relative magnitude  of  each of  the seven  main
effects and  the eight most important two-way  interaction effects (as
rows), for each luminosity bin  level (as columns).  The first column,
for example, shows  that over $95\%$ of the variance  in the number of
mock  bright  satellite  galaxies  in  the  V-band  magnitude  M$_{\rm
  v}=17.5$  is accounted  for  by variation  in  the baryon  fraction
$f_{\rm bary}$, with most of the remainder accounted for by variations
in  star  formation  efficiency  $\epsilon_*$. It  is  interesting  to
observe how,  as we move  toward fainter magnitudes, the  fraction of
variance explained  by variations of $f_{\rm  bary}$ decreases whereas
the one  associated with $z_{\rm  r}$ increases. The  transition takes
place at around  M$_{\rm v} \approx -11.5$.  At  this M$_{\rm v}$, the
interaction effect $z_{\rm  r} ~ : ~ f_{\rm  bary}$ becomes important,
indicating a coupling  of both input parameters. This  coupling can be
clearly observed on the top right panel of Figure~\ref{fig:maineff}.
 
From Figure~\ref{fig:fanova} we can infer that observables extracted
from the luminosity function could only be used to constrain
parameters such as $f_{\rm bary}$, $z_{\rm r}$ and, to a much lesser
extent, $\epsilon_{*}$.  In our models, the remaining four parameters
cannot account for significant variations of the cumulative number of
galaxies at any magnitude bin -- or, taken another way, the observables
we have chosen provide no meaningful constraints on these particular
parameters.  Thus, a different set of observables is required if we
wish to constrain any of the remaining model parameters.

In Figure~\ref{fig:maineff_feh} we show  the main effects computed for
mock observables extracted from the  metallicity function.  The first panel
shows  that   the  cumulative   number  of  satellite   galaxies  with
$\langle$[Fe/H]$\rangle  \geq  -1.1$ strongly  depends  on the  escape
factor  of  metals, $f_{\rm  esc}$.   There  is  also a  much  weaker
dependence  on the value  assigned to  the supernova  energy coupling,
$\epsilon_{\rm SN}$, and to the redshift of reionization, $z_{\rm r}$.
In a similar fashion to  what was observed for the luminosity function
observables,    as     we    move    toward     lower    values    of
$\langle$[Fe/H]$\rangle$  the  number  of satellite  galaxies  rapidly
increases   and   $z_{\rm   r}$   becomes  the   dominant   parameter.
Figure~\ref{fig:fanova_feh}    shows     the    corresponding    ANOVA
decomposition.   We  can  clearly  observe  how the  variance  on  the
cumulative number of metal-rich satellite galaxies (defined here to be
$\langle$[Fe/H]$\rangle  \geq  -1.1$)   is  closely  associated  with
variations of $f_{\rm esc}$  and only slightly on $\epsilon_{\rm sn}$,
whereas     the     cumulative     number    of     satellites     with
$\langle$[Fe/H]$\rangle  \geq  -2.2$  is  dominated by  the  parameter
$z_{\rm r}$.  The remaining parameters have a negligible effect on the
cumulative     number    of    galaxies     as    a     function    of
$\langle$[Fe/H]$\rangle$.  Interestingly, we observe a strong coupling
between  $z_{\rm  r}$  and  $f_{\rm  esc}$ at  almost  all  values  of
$\langle$[Fe/H]$\rangle$.     This   coupling    can   be    seen   in
Figure~\ref{fig:maineff_feh}.

The discussion in the previous paragraphs exemplifies the strengths of
the  ANOVA decomposition.  First,  it allows  us to  quickly determine
which  input   parameters  are  most  important   for  explaining  the
variability observed on a set  of model outputs.  Second, it allows us
to identify  which parameters \textbf{cannot}  be strongly constrained
by a given observational data set.   In our example both the SNII iron
yield, $m_{\rm Fe}^{\rm II}$, and  the SN Ia probability, $f_{Ia}$, do
not show a  strong influence on the model  output's variance, at least
within the ranges  in which we have allowed  these parameters to vary.
Thus, it  is possible to  reduce the dimensionality and  complexity of
our  problem by  fixing their  values to  some informed  prior.   As a
result of  this discovery,  in the work  that follows we  will discard
these parameters and work only with a five-dimensional input parameter
space.

It  is interesting  to repeat  this analysis  using  different dark matter-only
simulations to  explore how the  formation history of the  host galaxy
may affect our  results.  For this purpose we  have computed the ANOVA
decompositions of  model emulators  from training  sets obtained
after coupling ChemTreeN  with different galaxy formation histories.
In all  cases, the same design  was used to create  the training sets,
which consisted of  $n=500$ points.  In Figure~\ref{fig:anova_diff_dm}
we show the  results obtained by performing an  ANOVA decomposition on
the luminosity function outputs with  the simulations MW1 and MW2 (the
two remaining simulations yielded similar results, and thus we omit the
figures   showing   them  from   this   paper).   Interestingly,   the
decompositions show,  qualitatively, no substantial  difference in all
the  formation  histories   considered.  Although  the  fractions of
the variance explained by the different parameters may
slightly vary from one simulation  to another, the parameters that are
dominant  remain  the  same for all model outputs.   This
highlights an  important property of this kind  of analysis: \emph{The
  ANOVA  decomposition  allows  us  to  characterize  the  relationship
  between  the  input  parameters   and  the  desired  model  outputs,
  independently of  the corresponding  real observable values  and the
  underlying formation  history of our galactic  model.}  Note however
that, while  the ANOVA decompositions are equivalent  in all formation
histories, the actual values of  the model outputs can (and typically do) differ
from emulator to emulator.

Comparison of Figure~\ref{fig:fanova} and the left panel of
Figure~\ref{fig:anova_diff_dm} shows that the results were not altered
by either reducing the dimensionality of the problem or by considering
a different training set.  To test for convergence, we have used a
random sub-sample of $n=300$ training models to compute the ANOVA
decomposition of model MW1. The resulting decomposition showed no
significant differences with respect to that obtained with $n=500$
points. 

\section{Searching for best-fitting parameter regions in
  a multidimensional space}
\label{sec:fid}

\begin{figure*}
\centering
\includegraphics[width=180mm,clip]{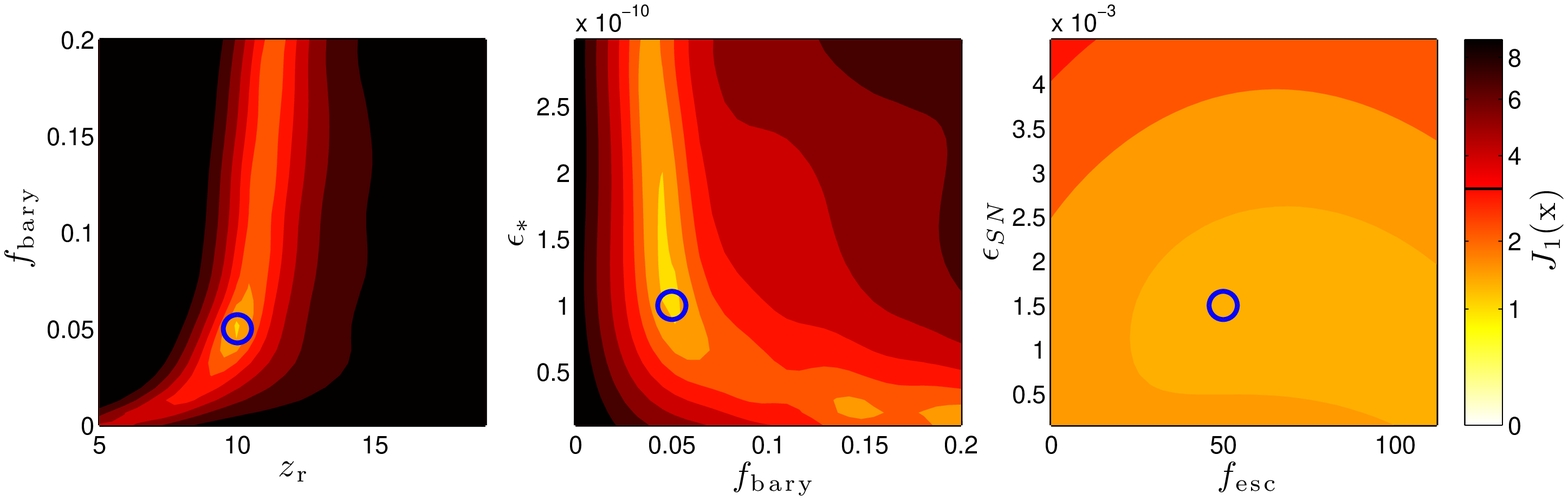}\\
\includegraphics[width=180mm,clip]{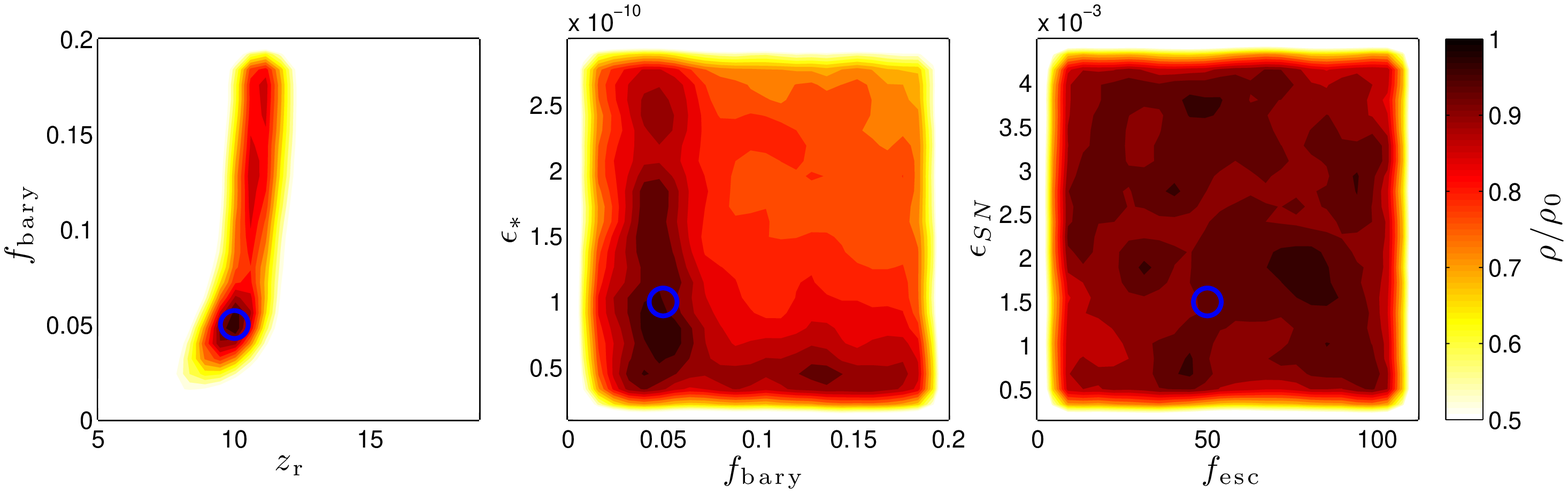}
\caption{Top  panels: Different sections  of the  joint implausibility
  surface, $J_{1}(\vx)$.   The different colors  show different values
  of $J_{1}(\vx)$ in logarithmic  scale.  Model emulators are compared
  to values  of the mock observable Luminosity  Function (LF) obtained
  after running  ChemTreeN with  the fiducial parameter  values.  Both
  the  mock observables  and the  training  data set  are obtained  by
  coupling  ChemTreeN with  the N-body  simulation MW1.   The fiducial
  values  of the  corresponding parameters  are indicated  with  a blue
  circle The  horizontal black solid  line on the color  bar indicates
  the imposed threshold: a value above this threshold shows that it is
  very implausible to obtain a good  fit to the observed data with the
  corresponding values  of the  model parameters.  Bottom  panels: Two
  dimensional projected densities of the DRAM chain points, obtained
  after marginalizing the samples  of the five dimensional likelihood,
  $\mathcal{L}_{1}(\vx)$ (see equation~\ref{eqn:likelihood}), over the
  remaining  three dimensions.   The  samples have  been smoothed  and
  contoured to aid the eye.   In all projections the posterior density
  has  been  normalized by  its  maximum  value.   The most  plausible
  regions of  input parameter space  are shown as the  highest density
  peaks.}
\label{fig:lf_chain}
\end{figure*}

\begin{figure}
\centering
\includegraphics[width=80mm,clip]{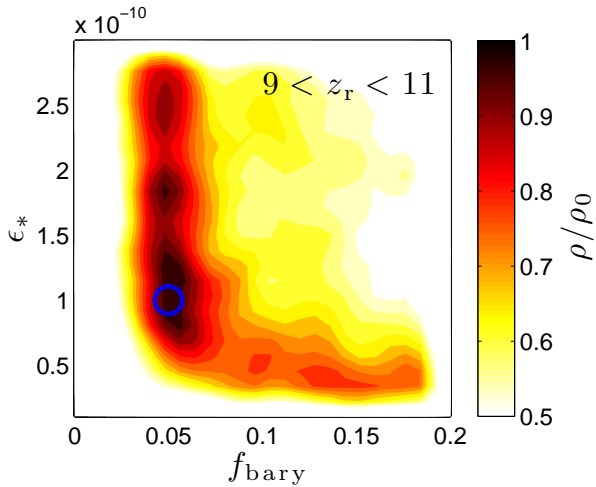}
\caption{As in  the bottom middle  panel of Figure~\ref{fig:lf_chain},
  obtained  after  only  considering  chain points  located  within  a
  restricted  range  of  $z_{\rm  r}$, centered  around  its  fiducial
  value. The corresponding range is  indicated in the top right corner
  of the panel.}
\label{fig:mcmc_red}
\end{figure}

\begin{figure*}
\centering
\includegraphics[width=180mm,clip]{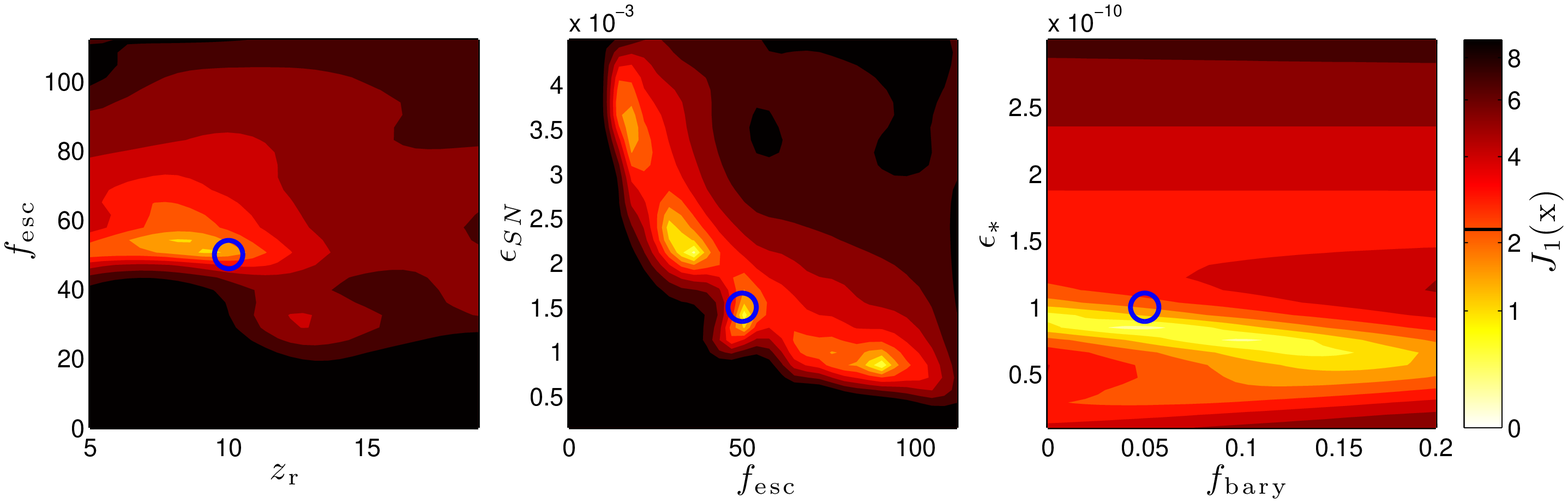}\\
\includegraphics[width=182mm,clip]{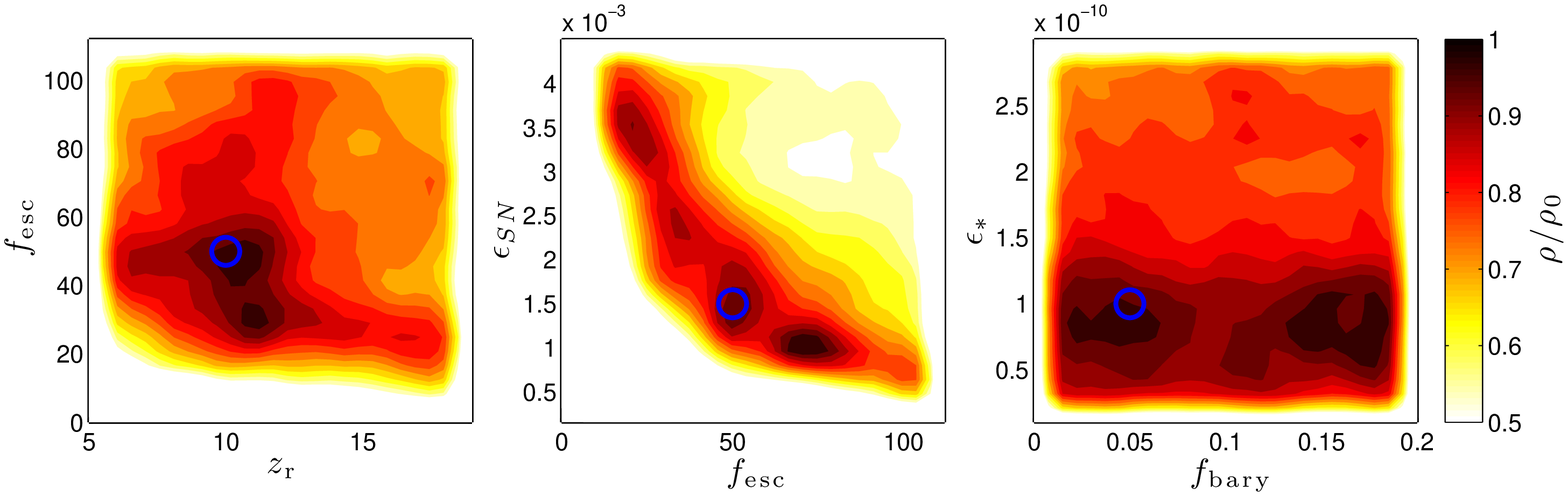}
\caption{As   in  Figure~\ref{fig:lf_chain},   when   values  of   the
  corresponding  Metallicity Functions  (MFs) are  considered  as mock
  observables. }
\label{fig:feh_chain}
\end{figure*}

  The  sensitivity  analysis  performed in  the  previous  section
  allowed us to  identify the set of input  parameters that, given the
  selected observables,  could be significantly  constrained.  In this
  section   we   introduce  a   technique   to  efficiently   identify
  best-fitting   input   parameter    regions   when   dealing   with
  multidimensional spaces, $p  > 3$. 

  To explore the robustness of the method we will first consider a set
  of  mock  observables obtained  after  coupling  ChemTreeN with  the
  simulation MW1.  The values of  the parameters used to generate this
  model,  and   those  we   will  try  to   recover,  are   listed  in
  \tabref{tab:params} as the fiducial model values.  These values were
  previously used by T10 to  fit reasonably well several properties of
  the  Galactic  stellar halo.   As  discussed  in  G12, best  fitting
  parameters could in principle be identified by searching for regions
  of  low values of  the joint  implausibility measure,  $J(\vx)$ (see
  Equation~(\ref{eqn-implaus-joint})).   In  low dimensionality  input
  parameter spaces,  $p \leq 3$, this  goal can be  achieved simply by
  slicing  the  resulting $J(\vx)$  data  cube.   However, for  higher
  dimensional  spaces  this  task  becomes  unfeasible.   Instead,  to
  explore the  input parameter space  we will use a  Delayed Rejection
  Adaptive Metropolis  (DRAM) sampling method.  DRAM is  the result of
  combining two powerful methods, i.e.  Delayed Rejection and Adaptive
  Metropolis,  to improve the  efficiency of  Metropolis-Hastings type
  Markov Chain Monte Carlo  (MCMC) algorithms \citep[for details about
  this method, see][]{dram}.

We start by constructing  Gaussian process model emulators considering
a five-dimensional input parameter  space that includes the parameters
$\vx        =       \left(z_{\rm        r},~f_{\rm       bary},~f_{\rm
    esc},~\epsilon_{*},~\epsilon_{\rm   SN}\right)$.    We  create   a
training set  consisting of $n=500$  points. Note that,  as previously
discussed, a  design with a smaller  number of points  could have been
considered.  However,  this relatively  large number of  design points
provides more  accurate emulators within  a reasonable run  time.  The
black  solid  lines  in  Figure~\ref{fig:hists}  show  the  cumulative
functions  extracted from our  fiducial model.   

The outputs from  our Gaussian process model emulators  and the set of
mock observables are used  to compute the joint implausibility measure
$J(\vx)$, shown  in Figure~\ref{fig:lf_chain} .  In the  top panels we
show  two-dimensional  sections of  the  $J_{1}(\vx)$ data  hypercube,
where the  sub index 1 indicates  that the training set  used to build
the  emulators was  obtained after  coupling ChemTreeN  with  the dark
matter  simulation MW1.  Note that  these sections  are the  result of
slicing the hypercube through  the \emph{known fiducial values} of the
remaining three  parameters.  In each  panel, the fiducial  values of
the two  remaining parameters  are indicated with  a blue  circle.  As
expected     from     the     ANOVA     decomposition     shown     in
Figure~\ref{fig:fanova},  with  this set  of  mock  observables it  is
possible to strongly constrain  the parameters $z_{\rm r}$ and $f_{\rm
  bary}$  (as  shown in  the  top left  panel).   Note  that the  most
plausible  regions enclose  the fiducial  values of  these parameters.
The  top  middle  panel  of Figure~\ref{fig:lf_chain}  shows  that  an
equally  good fit to  the luminosity  function can  be obtained  for a
large range  of $\epsilon_{*}$  values.  Clearly, constraints  on this
parameter are significantly weaker.   Note also that the remaining two
parameters,  $f_{\rm esc}$  and $\epsilon_{\rm  SN}$, are  very poorly
constrained (top right panel).

In reality, the values of the parameters that could best reproduce the
(real) observables are all  unknown.  As previously discussed, slicing
the resulting multidimensional data cube  to search for regions of low
$J(\vx)$ values becomes unfeasible for  values of $p > 3$.  To explore
the  input  parameter space  we  use  the  DRAM sampling  method.  The
likelihood used by the DRAM is
\begin{equation} 
\label{eqn:likelihood}
  \mathcal{L}_{1}(\vx) \propto e^{-J_{1}(\vx)^2 / 2},
\end{equation}
where we  have assumed a multivariate normal  distribution and uniform
prior for all parameters. Note that the assumed priors could be easily
modified  to  account  for  any  previous knowledge  about  the  input
parameters'    values.    Nonetheless,   as    we   show    later   in
Section~\ref{sec:realmw}, the choice of uniform priors is important if
we  want  to  characterize  the  dependence of  the  ``best  fitting''
parameter  selection process on  the merger  histories of  the adopted
Milky Way-like  models.   The resulting  joint  posterior distributions  are
shown  in the  bottom panels  of Figure~\ref{fig:lf_chain}.   The DRAM
chains  presented  in this  work  consist  of $5\times10^{5}$  points.
Convergence of  these chains was  assessed by diagnostics such  as the
Geweke  test  \citep{geweke}.  Each  panel presents  contours  of  the
projected  density of points,  $\rho$, obtained  from the  DRAM chain.
The  two-dimensional  projected  densities  represent  the  result  of
marginalizing  the  DRAM  chain   samples  over  the  remaining  three
dimensions.  For comparison, in all cases we have normalized $\rho$ to
it  maximum value, $\rho_{0}$.   The most  plausible regions  of input
parameter space are shown as the highest density peaks.  Note that the
fiducial  values of the  parameters $z_{\rm  r}$, $f_{\rm  bary}$, and
$\epsilon_{*}$  are located  within the  highest density  regions.  As
expected, however,  the parameters $z_{\rm r}$ and  $f_{\rm bary}$ are
significantly  more  strongly  constrained  than  $\epsilon_{*}$.   To
explore whether spurious structure  in these density contours could be
induced due to auto-correlation in  the chain, we have split the chain
into  five different ``subchains''.   This thinning  of the  chain was
done by taking one out of every five points, with a different starting
point taken from  the first five elements of the  total chain.  In all
cases,   the  results   were   not  affected   by  this   sub-sampling
\citep[see][for an interesting discussion on thinning]{thinning}.

In general, the shape of the density contours is very similar to that
of the $J_{1}(\vx)$ sections, shown on the top panels. \emph{It is
important to note that the DRAM chain density contours are obtained
after fully sampling the five-dimensional input parameter space,
without any prior knowledge of the fiducial values of the parameters}.
To obtain these two-dimensional density contours we are implicitly
averaging over all the variations in the three remaining directions.
Instead, to compute the two-dimensional $J_{1}(\vx)$ sections, the
$J_{1}(\vx)$ hypercube was sliced at the fiducial values of the
remaining three parameters.  Thus, prior knowledge of these
parameters' values was required. A similar idea can be applied to the
DRAM chain density to improve the constraint on the star formation
efficiency, $\epsilon_{*}$.  As previously discussed, the bottom left
panel of Figure~\ref{fig:lf_chain} imposes strong constraints on the
redshift of the epoch of reionization, $z_{r}$.  In
Figure~\ref{fig:mcmc_red} we show DRAM density contours in $f_{\rm
  bary}$ and $\epsilon_{*}$ space, obtained after only considering
chain points located within a restricted range of $z_{\rm r}$,
centered around its fiducial value. The chosen range, $9 < z_{\rm r} <
11$, is large enough to fully include the high density region shown on
the bottom left panel of Figure~\ref{fig:lf_chain}.  Note that as a
result of choosing this reasonable range in $z_{\rm r}$, both $f_{\rm bary}$ and $\epsilon_{*}$ are significantly
better constrained.

In Figure~\ref{fig:feh_chain} we show $J_{1}(\vx)$ sections (top row),
and the corresponding DRAM chain density contours (bottom row),
obtained when values of the metallicity function are considered as
mock observables.  The ANOVA decomposition shown in
Figure~\ref{fig:fanova_feh} indicated that, in our models, a
significant fraction of the variability observed in these model
outputs is associated with variations of the parameters $z_{\rm r}$,
$f_{\rm esc}$, and $\epsilon_{\rm SN}$. Indeed, the $J_{1}(\vx)$
sections show that constraints to these parameters can be obtained
when these mock observables are considered. This is especially relevant for
the pair of parameters $f_{\rm esc}$ and $\epsilon_{\rm SN}$, which
could not be constrained by the luminosity function. Note that the
corresponding section (middle panel) unveils a non-linear relation
between these two parameters, with several ``islands'' of very low
implausibility (i.e., high probability).  In each section, the
fiducial values of corresponding pairs of parameters are indicated with
a blue circle.  As previously shown for the luminosity function, these
parameters can be significantly constrained without any prior
knowledge of the parameters' fiducial values thanks to the DRAM
chain sampling.

\begin{figure}
\centering
\includegraphics[width=80mm,clip]{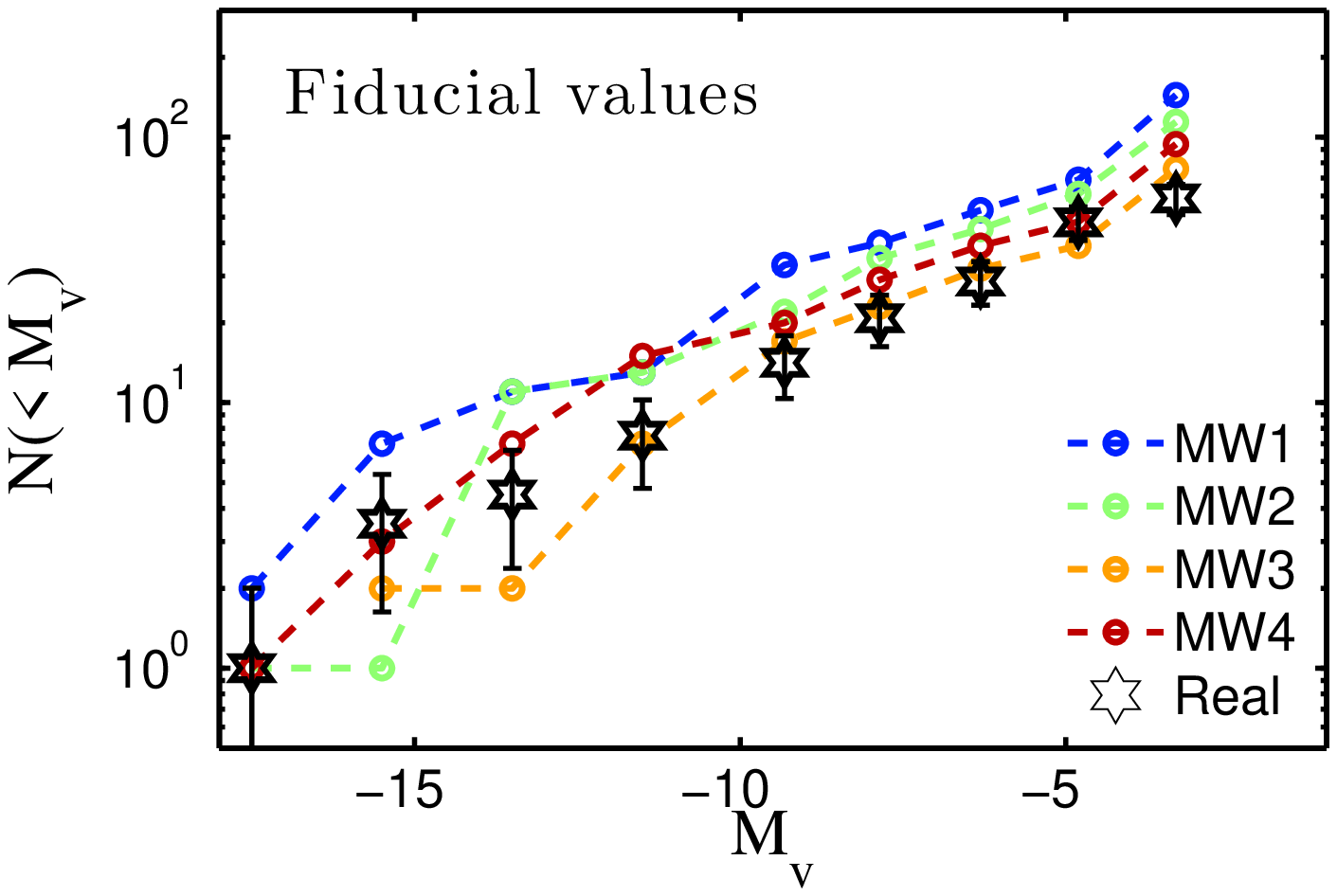}\\
\includegraphics[width=80mm,clip]{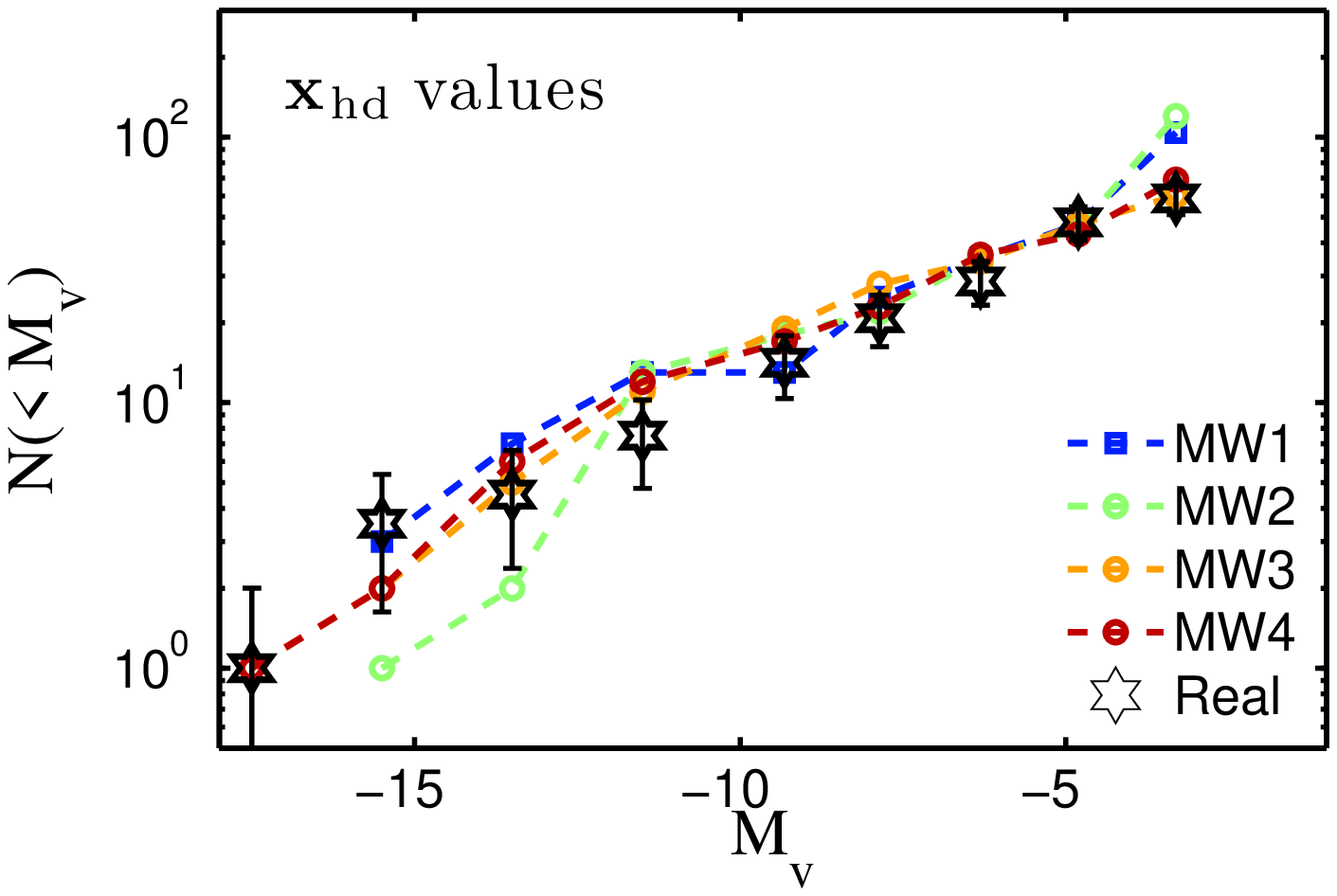}
\caption{Cumulative  number of  satellite  galaxies as  a function  of
  absolute  V-band magnitude,  M$_{\rm  v}$. The  top  panel shows  the
  results obtained when the input parameters are fixed to the fiducial
  values,  listed in  \tabref{tab:params}. The  bottom panel  shows the
  results obtained  when the input  parameters are fixed  at $\vx_{\rm
    hd}^{i}$, the highest density peak of the corresponding  DRAM chain
  (see \tabref{tab:params_xhd}). In both panels, the black stars show
  the luminosity function of observed Milky Way satellite galaxies corrected for incompleteness
  as described by \citet{2008ApJ...686..279K}. The bars indicate
 Poisson noise error.}
\label{fig:lf_t10}
\end{figure}

\section{Applications to the Milky Way: A method to constrain its
  assembly history}
\label{sec:realmw}

\begin{figure*}
\centering
\includegraphics[width=57mm,clip]{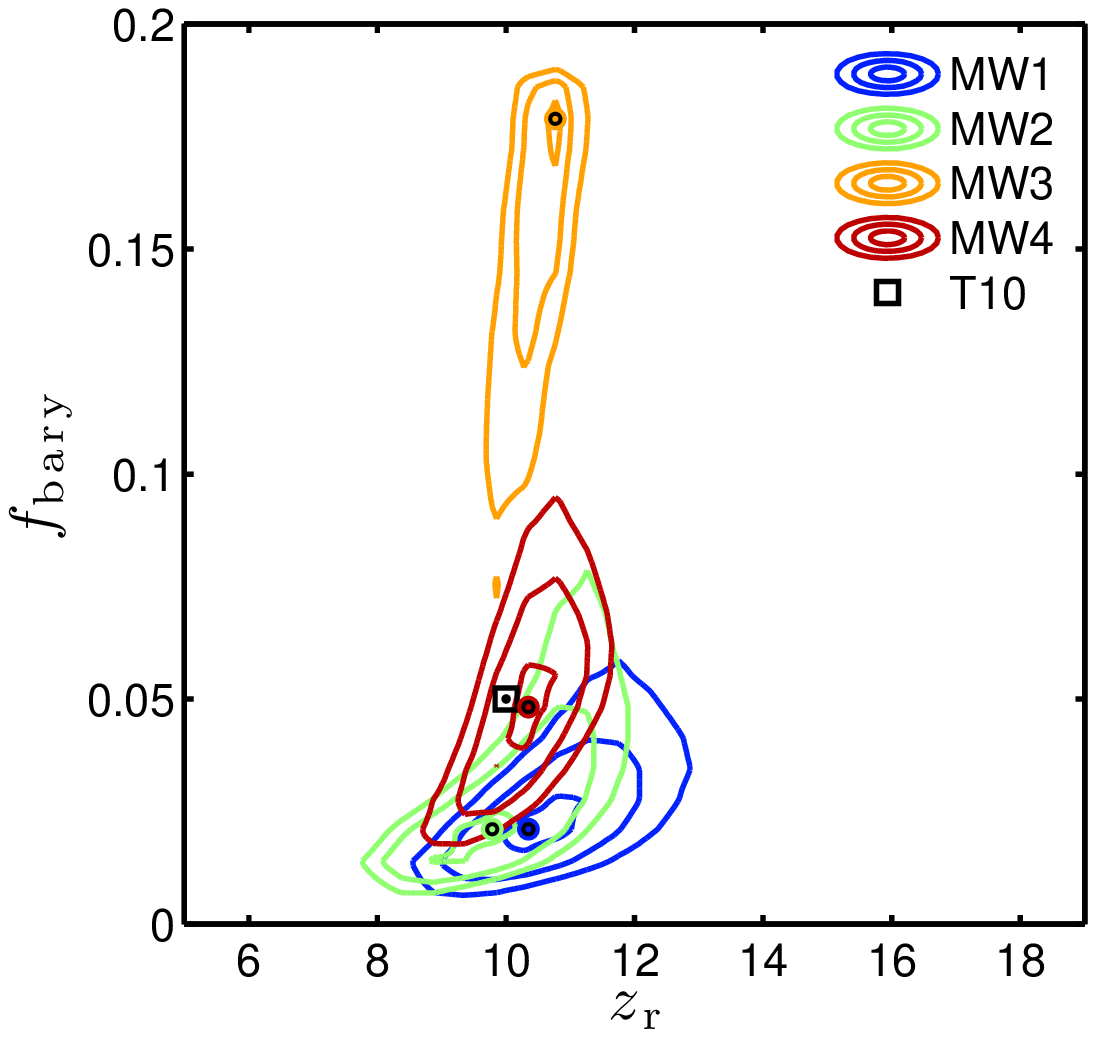}
\includegraphics[width=57mm,clip]{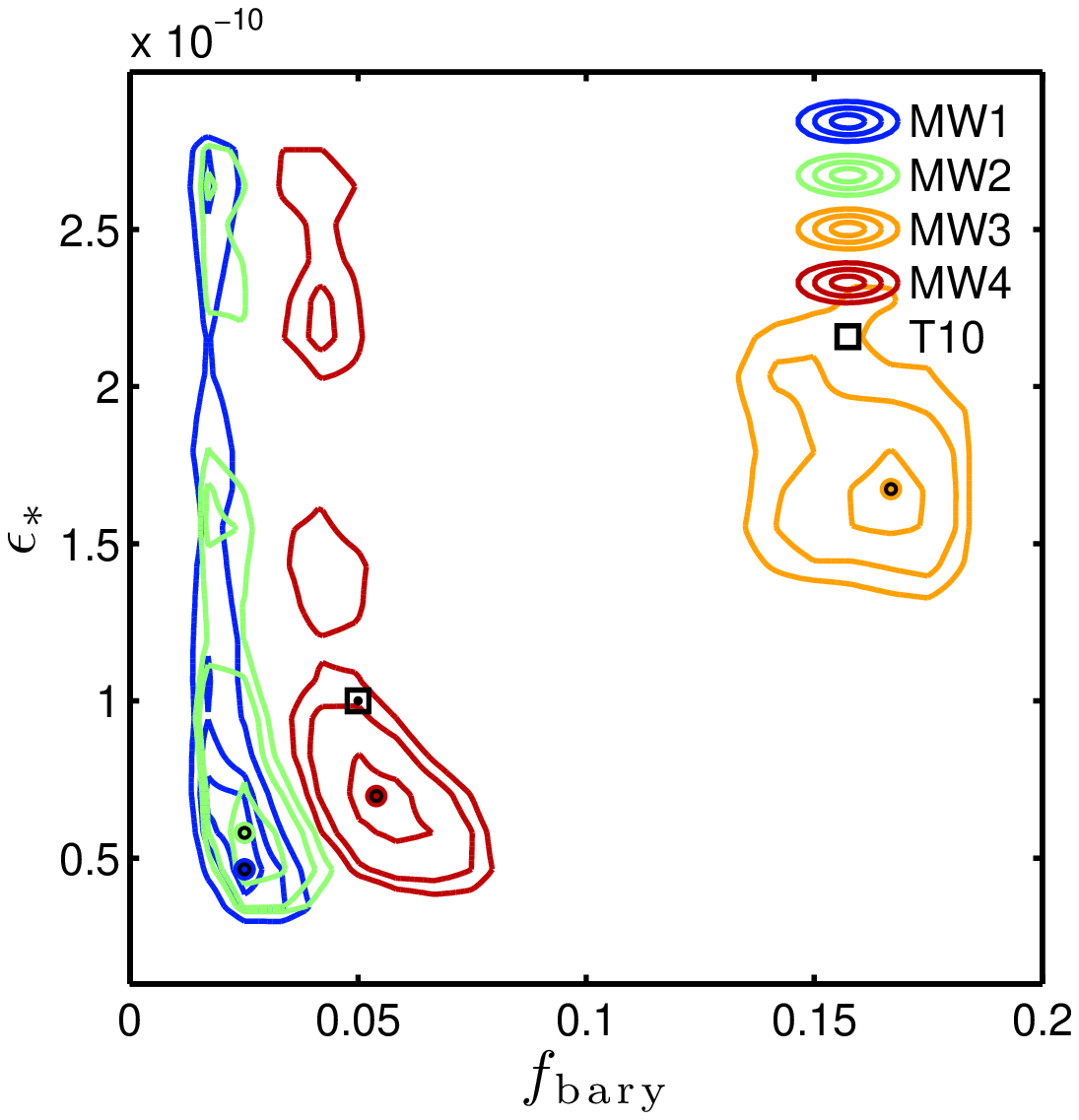}
\includegraphics[width=57mm,clip]{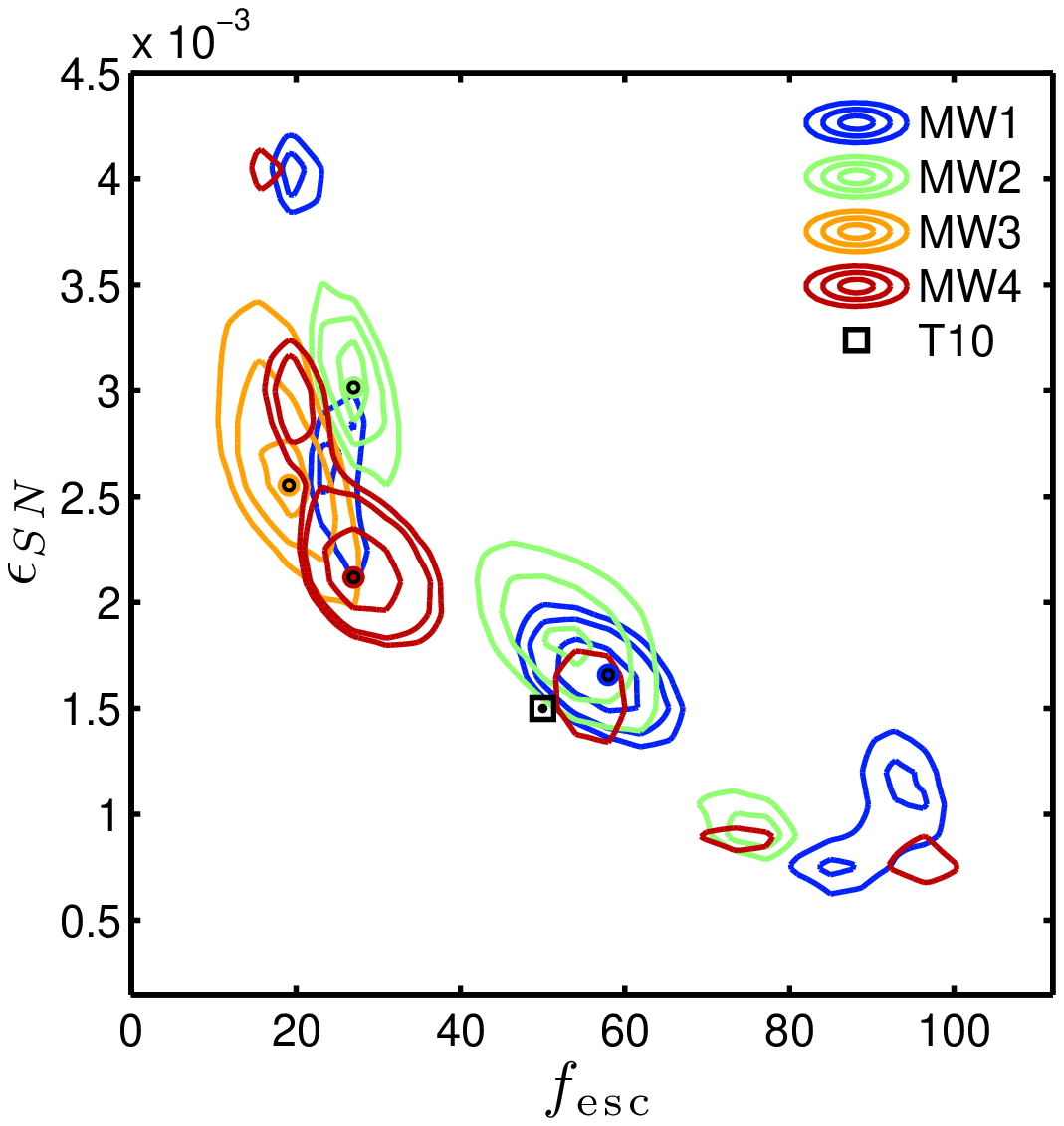}
\caption{Projected densities of DRAM chain points obtained from our
different MW$i$ models, with $i=1,2,3$ and 4.  The color-coded
contours show the results obtained with a different likelihood
function, $\mathcal{L}_{i}(\vx)$.  Starting from the densest point of
each final distribution, $\vx^{i}_{\rm hd}$, the different contour
levels enclose 1, 5 and 10 per cent of corresponding DRAM-chain
points.  The color-coded dot indicates the location of $\vx^{i}_{\rm hd}$, whereas the black dots indicate the location of the parameter's
fiducial values. The black square indicates the `best-fit' model from 
\citet{tumlinson2}.}
\label{fig:mcmc_lf}
\end{figure*}

 In G12 we showed that the best-fitting input parameter selection
  process strongly  depends on the underlying merger  history of Milky
  Way-like galaxy used to train  the model emulators.  In this section
  we will show how this characteristic  of our method could be used to
  constrain the assembly  history of the Milky Way  and its properties
  at $z=0$. Our approach consists of obtaining a best-fitting model for
  each Milky Way-like dark matter halo being considered.  The best-fit
  parameters  are allowed  to freely  vary from  model to  model.  The
  resulting  best-fit  model  is  then  compared  with  a  second  and
  independent observational data set that can then be used to evaluate
  its reasonableness.  As we will  show in what follows, it is always
  possible to  find a set of  input parameters to  tightly reproduce a
  given observational  data set. However, only some  of these best-fit
  models are successful at reproducing a second and independent set of
  observables. The results from  the sensitivity analysis presented in
  Section~\ref{sec:in-out} will allow us  to focus our analysis on the
  parameters that  have the largest  impact on the predictions  of the
  selected  observables. The  DRAM  method will  allow  us to  quickly
  explore  the resulting joint  implausibility hypercubes  to identify
  regions of best fitting input parameter sets.

The observables that are considered in this Section are extracted from
the  real Milky Way  galaxy's satellite  distribution.  These  are the
luminosity  function of  satellite  galaxies located  within 280  kpc,
corrected      for       incompleteness      as      described      by
\citet{2008ApJ...686..279K}, and  the cumulative metallicity function.
Values for  the $\langle$[Fe/H]$\rangle$  of all dwarfs  were extracted
from  the data  compilation presented  by \citet{2012AJ....144....4M}.
Note that,  thus far, the fiducial  model and the  training set shared
the same  galaxy's formation  history, associated with  the simulation
MW1. The formation history of the Milky Way is of course an unknown in
our search for the best-fitting parameters.

Figure~\ref{fig:lf_t10}  shows,  with   black  stars,  the  luminosity
function     of     the     Milky     Way's     satellite     galaxies
\citep{2012AJ....144....4M}.   For comparison, the  color-coded dashed
lines show  the luminosity  function of the  models with  the fiducial
parameters.  These models were  obtained after coupling ChemTreeN with
the four dark matter-only cosmological simulations MW$i$ (with $i = 1,
2, 3$ and  4), and fixing the input parameters  at the fiducial values
listed  in  \tabref{tab:params}.   The  four models  show  significant
deviations from the  data.  It is thus likely that  for each Milky Way
dark matter halo model there exists a small volume of input parameter
space within  which a better  fit to the observed  luminosity function
can  be  obtained.  To  search  for this  volume  we  employ the  DRAM
sampling  technique previously  described.  We  train  model emulators
using  four different  training sets.   The  sets are  the results  of
coupling   ChemTreeN  to  the   four  dark   matter-only  cosmological
simulations  of Milky  Way-size  galaxies. The  same  design for  each
simulation, consisting of  $n=500$ points, was used.  For  each set of
model emulators, trained on a different MW$i$, we obtained a different
joint       implausibility       function      $J_{i}(\vx)$       (see
eqn~(\ref{eqn-implaus-joint})).  These $J_{i}(\vx)$  are the result of
comparing the  outputs of the  model emulators to the  real observable
data.  We use the  $J_{i}(\vx)$ to construct four different likelihood
functions                  $\mathcal{L}_{i}(\vx)$                 (see
eqn.~\ref{eqn:likelihood}). Figure~\ref{fig:mcmc_lf} shows the results
of the DRAM  sampling.  The left panel show  contours of the projected
density of  DRAM chain  points in $(z_{\rm  r}, f_{\rm  bary})$ space.
The different colored contours shows the results obtained with the four
different $\mathcal{L}_{i}(\vx)$.  Starting  from the densest point of
each  final distribution,  $\vx^{i}_{\rm hd}$,  the  different contour
levels  enclose 1,  5  and 10  per  cent of  corresponding DRAM  chain
points.  The  color-coded dot indicates the  location of $\vx^{i}_{\rm
  hd}$.  Note  that strong constraints on the  parameters $(z_{\rm r},
f_{\rm bary})$  are obtained for the  four MW$i$.  Let  us recall that
the satellite  luminosity function is  most sensitive to this  pair of
input  parameters.   Interestingly,  except  for the  model  MW4,  the
locations  of  $\vx^{i}_{\rm  hd}$  are  significantly  off  from  the
fiducial values,  especially in the direction of  $f_{\rm bary}$.  The
values of the parameters associated with $\vx^{i}_{\rm hd}$ are listed
in \tabref{tab:params_xhd}.   The most extreme  case is given  by halo
MW3, where the most plausible value of $f_{\rm bary}$ is approximately
four times larger than the fiducial value.  Note that, as shown in the
top panel of Figure~\ref{fig:lf_t10}, when compared with the Milky Way
luminosity function this model (obtained with the fiducial parameters)
presents  a significant deficit  of bright  satellites.  On  the other
hand, MW1's  model shows  an excess of  satellites at  all magnitudes.
Note that the  most plausible value of $f_{\rm  bary}$ obtained by the
DRAM sampling in  this case is approximately two  times lower than the
fiducial value.  To explore whether the location of $\vx^{i}_{\rm hd}$
depends on the number of points  used in the DRAM sampling, we divided
the  final  chains  into  five  different subchains  as  described  in
Section~\ref{sec:fid}.    From   each   sub-chain  we   obtained   the
corresponding location of $\vx^{i}_{\rm  hd}$ and computed its average
value,  $\langle \vx^{i}_{\rm  hd}  \rangle$.  We  find  that, in  all
cases, $\langle  \vx^{i}_{\rm hd}  \rangle$ is in  excellent agreement
with the  values of $\vx^{i}_{\rm  hd}$ obtained from the  full chain.
In  most  cases  the  associated  standard  deviation  is  negligible.
Furthermore, as we will show  below our results are not significantly
affected by small variations  in $\vx^{i}_{\rm hd}$.  The middle panel
of Figure~\ref{fig:mcmc_lf} shows contours of the projected density of
DRAM-chain  points  in   the  $(f_{\rm  bary},  \epsilon_{*})$  space.
Following  the discussion  in Section~\ref{sec:fid},  to  obtain these
contours we  only considered  chain points that  are located  within a
specific range  of $z_{\rm r}$,  centered around the  value associated
with $\vx^{i}_{\rm  hd}$.  The  range of $z_{\rm  r}$ chosen  for each
MW${i}$ is  such that  it includes all  chain points that  are located
within the regions defined by the 10 per cent contour levels, shown in
the   left   panel   of  Figure~\ref{fig:mcmc_lf}.    Constraints   on
$\epsilon_{*}$ are weaker  than those found for the  pair $(z_{\rm r},
f_{\rm  bary})$.  Multiple  plausible regions  of parameter  space are
found for almost all MW$i$  .  The values of $\epsilon_{*}$ associated
with $\vx^{i}_{\rm hd}$  show a large scatter and,  as before, it gets
closer to the fiducial value for MW4.

\begin{center}
\begin{deluxetable}{lcccccc}
  \tabletypesize{\footnotesize} \tablecaption{Model parameter extracted from the highest density peak of the
    corresponding DRAM-chain's posterior density.  \label{tab:params_xhd}}
  \tablewidth{240pt}

  \tablehead{\colhead{Name}& \colhead{$z_{\rm r}$} & \colhead{$f_{\rm esc}$} & \colhead{$f_{\rm bary}$} &
    \colhead{$\epsilon_{*}$} & \colhead{$\epsilon_{\rm SN}$} & \colhead{$M_{40}$\tablenotemark{a}}}
  \startdata
  \MW1 & 10.3 & 57.7 & 0.021 & 0.4 $\times 10^{-10}$  & 0.00165 & 0.57  \\
  \MW2 & 9.6 & 27.0 & 0.021 &  0.5 $\times 10^{-10}$ & 0.00304 & 0.95 \\
  \MW3 & 10.7 & 19.1 & 0.168 & 17 $\times 10^{-10}$ & 0.00255 &  20.9 \\
  \MW4 & 10.3 & 27.0 & 0.048 & 0.7 $\times 10^{-10}$ & 0.00211 & 1.88
\enddata
\tablenotetext{a}{Masses are listed in $10^{8}~M_{\odot}$}
\end{deluxetable} 
\end{center}

\begin{figure}
\centering
\includegraphics[width=80mm,clip]{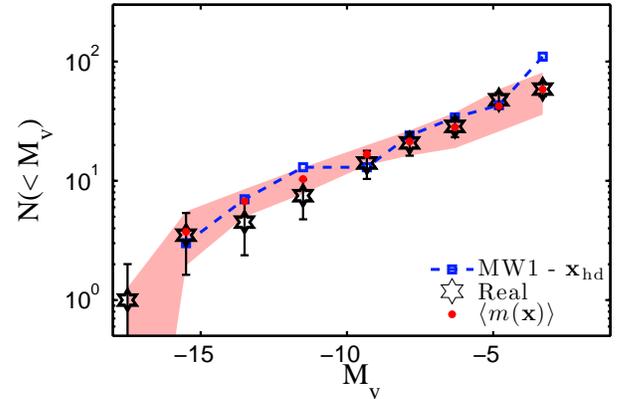}
\caption{Cumulative  number of  satellite  galaxies as  a function  of
  absolute  V-band magnitude, M$_{\rm  v}$. The  black stars  show the
  luminosity  function  of   observed  Milky  Way  satellite  galaxies
  corrected      for      incompleteness      as     described      by
  \citet{2008ApJ...686..279K}.  The bars indicate Poisson errors.
  The blue dashed  line shows, for model MW1,  the luminosity function
  obtained when the input  parameters are fixed at $\vx_{\rm hd}^{i}$,
  the  highest density  peak of  the corresponding  DRAM  chain (see
  \tabref{tab:params_xhd}).   The red  dots show  the  mean luminosity
  function  obtained  after  averaging  the prediction  of  the  model
  emulator for all DRAM points within  the 5 per cent contour shown in
  Figure~\ref{fig:mcmc_lf}. The red  shaded area show the $95\%$
  confidence interval.}
\label{fig:emu_mean}
\end{figure}

The  bottom  panel  of  Figure~\ref{fig:lf_t10} shows  the  luminosity
functions obtained after fixing the  values of the input parameters at
$\vx^{i}_{\rm hd}$.  The values  of $(f_{\rm esc}, \epsilon_{\rm SN})$
are kept fixed  at the fiducial values, as  the luminosity function is
insensitive     to    variation     of    these     parameters    (see
Figures~\ref{fig:fanova}  and \ref{fig:anova_diff_dm}).   It  is clear
that,  in all  cases, a  much better  fit to  the  observed luminosity
function is obtained  with the sets of most-likely parameters derived
from our DRAM chains.  We now  explore how sensitive this result is to
the exact  location of $\vx^{i}_{\rm  hd}$.  For model MW1,  we select
the $\approx  2.5 \times 10^{4}$  DRAM points that are  located within
the 5 per cent contour shown in Figure~\ref{fig:mcmc_lf}, and obtain a
predicted  luminosity function for  each these  points.  Note  that no
restrictions are applied to the parameters $(\epsilon_{*}, f_{\rm esc}
, \epsilon_{\rm SN})$.  The  resulting luminosity function computed by
averaging all these points  is shown in Figure~\ref{fig:emu_mean} with
red  dots.  The  red  shaded  area  indicates  the  $95\%$  confidence
interval.  A very good fit  to the observed luminosity function is also
obtained in  this case, indicating  that our results are  not strongly
sensitive to the exact location  of $\vx^{i}_{\rm hd}$.  

\begin{figure}
\centering
\includegraphics[width=78mm,clip]{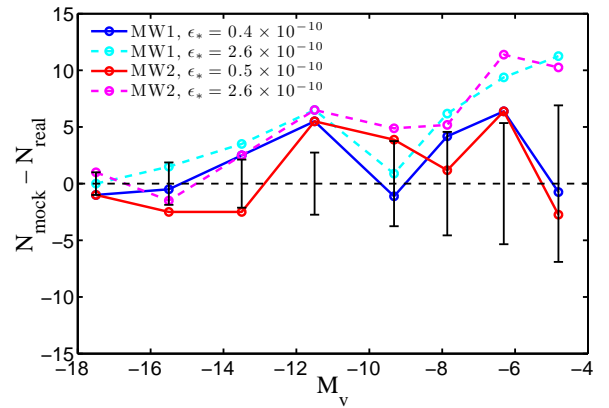}
\caption{Residual luminosity functions obtained after subtracting the
  real  luminosity function from  the results  of different  models. The
  solid lines  indicate the results obtained when  the value of
  $\epsilon_{*}$  associated with  $\vx_{\rm hd}$  is  considered. The
  dashed   lines  show  the  results obtained  when  the  largest
  plausible value  of $\epsilon_{*}$ is considered.  The horizontal black dashed
  line   indicates  a   perfect  fit   to  the   observed  luminosity
  function. The errors bars indicate Poisson errors.}
\label{fig:lf_comp}
\end{figure}

A good  fit to the luminosity  function, however, does  not imply that
the four  resulting models, associated with the  different MW$i$s, are
equally  good  at   reproducing  simultaneously  both  the  luminosity
function of Milky Way dwarf satellites and the properties of the Milky
Way  stellar  halo.   As  an  example,  we compare  the  mass  of  the
corresponding  stellar  halos within  $1-40$  kpc,  $M_{40}$ with  its
observationally-determined value  for the Milky Way.   Using the Sloan
Digital Sky Survey (SDSS),  \citet{bell08} estimated a mass of $M_{40}
= (3.7 \pm 1.2) \times 10^8~M_{\odot}$ for the Milky Way stellar halo.
The values of  $M_{40}$ in our four best-fitting  models are listed in
\tabref{tab:params_xhd}\footnote{The listed values have been corrected
  by   the  different   mass-to-number  stellar   ratios   adopted  by
  \citet{bell08}  and \citet{tumlinson2}}.  Interestingly,  models MW1
and  MW2 present  an $M_{40}$  that is  significantly smaller  than the
observationally-determined value,  whereas MW3 suggests  a much larger
value.   The simulated  $M_{40}$  is comparable  to its  observational
counterpart  only for  MW4.  Note  that  this result  even holds  when
fixing   the   value   of   the   less   well-constrained   parameter,
$\epsilon_{*}$, to  its fiducial value.   Models MW1, MW2 and  MW3 are
thus less  likely to represent a good  model of the Milky  Way and its
underlying  formation history  than MW4.   Nonetheless,  as previously
discussed, constraints  on $\epsilon_{*}$  are poor and  thus multiple
high density  regions with different  values of this parameter  can be
seen  in  the  middle  panel  of  Figure~\ref{fig:mcmc_lf}.   This  is
especially true for models MW1 and MW2. As an example, we consider for
these  two  models the  high  density  peak  located at  $\epsilon_{*}
\approx 2.6 \times 10^{-10}$.  This  is the largest plausible value of
$\epsilon_{*}$  for both  models.  The  modeled $M_{40}$  obtained are
$2.4$ and $3$ $\times 10^{8}~M_{\odot}$ for MW1 and MW2, respectively.
Whereas model MW1  cannot match the observed $M_{40}  $ even with this
extreme value of $\epsilon_{*}$, model  MW2 shows a better match.  The
luminosity functions  obtained with both, i.e.,  the largest plausible
value and that  associated with $\vx_{\rm hd}$, show  good fits to the
observed  luminosity  function.   However,  the former  results  in  a
slightly poorer fit.  This is shown in Figure~\ref{fig:lf_comp}, where
we plot  the residuals  of the luminosity  functions, $N_{\rm  mock} -
N_{\rm  real}$, for  both  values of  $\epsilon_{*}$.   Note that,  in
general, the luminosity functions associated with the larger plausible
values  of $\epsilon_{*}$  tend to  overpredict the  number  of faint
satellite galaxies. 

The analysis just performed has the potential to allow us to constrain
the Milky  Way's formation  history and its  properties at  $z=0$.  As
discussed in  Section~\ref{sec:sims}, the four halos  analyzed in this
work were specially targeted to resemble the Milky Way.  That is, they
all have a very similar virial  masses and have not experienced a
major  merger after  approximately $z  =  1.5-2$.  Their  growth as  a
function of time is shown in  Figure 4 of G12. Some differences can be
easily observed.  For example,  our best-fitting model associated with
halo MW4 has experienced the most significant late accretion. However,
our sample of Milky Way-like dark  matter halos is very small and thus
we are  strongly undersampling the range of  possible merger histories
of the Milky Way-like candidates.  A much larger sample is required to
determine  whether  any  particular  features  observed  in  a  halo's
assembly history are statistically significant.  Nonetheless, using the
set    of    simulations     from   the    Aquarius    project
\citep{2008MNRAS.391.1685S},   \citet{2013MNRAS.429..725S}  (hereafter
S13) find that the number of luminous satellite galaxies brighter than
$M_{\rm v}  = -5$ within the virial  radius of the  host shows a
significant correlation with  the host's dark matter halo virial
mass.  As discussed  by S13, \citet{2010MNRAS.402.1995M} also observed
this  trend and remarked  that it  does not  depend on  the particular
semi-analytical   model  used.   As   shown  in   the  top   panel  of
Figure~\ref{fig:lf_t10},  the   same  behavior  is   observed  in  our
simulations.  Let  us recall  that  this  panel  shows the  luminosity
functions of our four Milky  Way-like models obtained after fixing the
input  parameters  to   their  fiducial  values.   Interestingly,  our
preferred best fitting model,  MW4, which can simultaneously reproduce
the number of  bright satellites with $M_{\rm v}  \leq -5$ (see bottom
panel  of  Figure~\ref{fig:lf_t10}),  and  can  provide  a  reasonable
estimate  of $M_{40}$,  has  a value  of  $M_{\rm vir}  = 1.44  \times
10^{12}$  $M_{\odot}$. This  value is  in good  agreement  with recent
estimates  of  the total  Milky  Way mass  \citep{2010ApJ...720L.108G,
  2013ApJ...768..140B,    2013ApJ...764..161K,   2013arXiv1309.4293P}.
Within our  framework, lower  mass halos such  as MW3, with  an $M_{\rm
  vir} = 1.22 \times 10^{12}$ $M_{\odot}$, would be ruled out.

\begin{figure}
\centering
\includegraphics[width=78mm,clip]{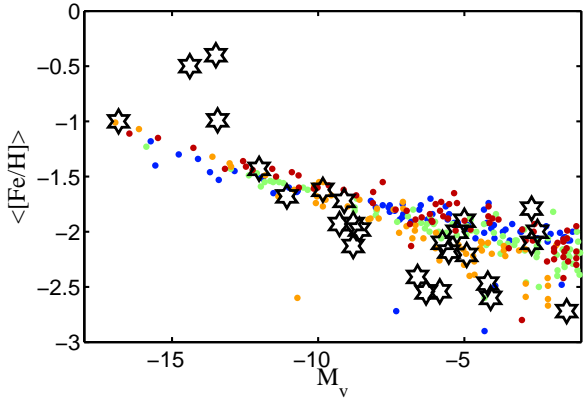}
\caption{Luminosity-metallicity ({\it L-Z}) relation of satellite
galaxies.  The black stars show the Milky Way {\it L-Z} relation, as
presented by \citet{2012AJ....144....4M}.  The color-coded dots show
the results obtained from our four models after fixing ChemTreeN's
input parameters at $\vx_{\rm hd}^{i}$, the highest-density peak of
the corresponding DRAM chain (see \tabref{tab:params_xhd})}
\label{fig:mv_feh}
\end{figure}

In   Figure~\ref{fig:mv_feh}   we   show,   with  black   stars,   the
luminosity-metallicity ({\it  L-Z}) relation of  Milky Way satellites.
Due  to incompleteness  in our  current sample  of  observed satellite
galaxies, and  uncertainties on $\langle$[Fe/H]$\rangle$ measurements,
it is not possible to  derive a complete metallicity function relation
down  to  $\langle$[Fe/H]$\rangle  =  -2.2$. The  sample  is  severely
incomplete  at  the metal-poor  (and  faint)  end  of the  metallicity
function.   Thus, in order  to compare  with our  models, we  derive a
metallicity function only taking  into account satellite galaxies more
metal-rich  than $\langle$[Fe/H]$\rangle  \geq -1.5$.   This  limit on
$\langle$[Fe/H]$\rangle$ imposes  a limit on M$_{\rm  v} \lesssim -11$
(see Figure~\ref{fig:mv_feh}), i.e., within the realm of the classical
dwarfs. Following \citet{toll08}, we assume that all satellites within
this magnitude range should have  been discovered anywhere in the sky,
with the possible exception of objects at low Galactic latitudes where
Milky   Way   extinction    and   contamination   become   significant
\citep{will04}.

In   practice,  we  train  a   model  emulator  
considering  only  two    bins   of    the    metallicity   function,    at
$\langle$[Fe/H]$\rangle  = -1.1$  and  -1.5.  As  shown  by the  ANOVA
decomposition in Figure~\ref{fig:fanova_feh}, the number of satellites
in these bins is very sensitive to variations of the escape factor of
metal, $f_{\rm  esc}$.  Variation of  the remaining parameters  does not
account for a  very significant fraction of the  variance obtained for
these observables.  The most metal-poor bins that we are
omitting from this  analysis are very sensitive to  the redshift of the
epoch  of  reionization,  $z_{\rm  r}$.   The lack  of  the  additional
constraints  provided  by  these  metal-poor  bins  could  induce  the
detection of spurious high density peaks in the DRAM density contours
associated with values  of $z_{\rm r}$ that do  not represent the real
data. To avoid this, we update the range of the uniform prior assigned
to $z_{r}$ in our DRAM sampling analysis. This update is done based on
the  results  obtained from  the  independent  observational data  set
associated with the luminosity function.   The new range for our $z_{\rm
  r}$  uniform priors  is such  that, in  all cases,  it  includes the
region of $z_{\rm  r}$ enclosed within the 10  percent contours shown
in the left panel of  Figure~\ref{fig:mcmc_lf}.  In the right panel of
the  same figure we  show contours  of the  projected density  of DRAM
chain points  in the $(f_{\rm esc}  , \epsilon_{\rm SN}  )$ space.  As
previously seen in  Figure~\ref{fig:fanova_feh}, a non-linear relation
with  several high  density  peaks  regions is  obtained  in all  four
models.  For a given value  of $\epsilon_{\rm SN}$, models MW3 and MW4
require a lower  value of $f{\rm esc}$ than models MW1  and MW2 to fit
the  observed  metallicity  function.   The values  of  the  parameter
associated  with the  highest  density peak,  $\vx_{\rm hd}^{i}$,  are
indicated  with colored  dots and  listed  in \tabref{tab:params_xhd}.
However, we remind  the reader that these values  should be taken with
caution due  to the large uncertainties in  the observable quantities.
In  Figure~\ref{fig:mv_feh} we  show  the {\it  L-Z}  relation of  the
models  obtained  after  fixing  the  input  parameters  at  $\vx_{\rm
  hd}^{i}$. Note that,  in all cases, a very good  fit to the observed
{\it L-Z} relation is obtained.

\section{Discussion and Conclusions}
\label{sec:conclusions}

  In this  paper  we have  presented  a novel application of  the
  statistical tool known as sensitivity analysis to characterize the
  relationship between input  parameters and observational predictions
  for  the  chemo-dynamical  galaxy  formation model  ChemTreeN.   In
particular, we  focus on efforts to  model the Milky  Way stellar halo
and   its  population   of   satellite  galaxies.    ChemTreeN  is   a
semi-analytic  model of  galaxy  formation that  has  been coupled  to
cosmological simulations  that provide realistic  merger histories and
phase space distribution of the resulting stellar populations.

The implementation  of a semi-analytic model  involves the fine-tuning
of a large number of free parameters that control the behavior of many
different  physical  processes,  and  the  choice of  a  ``best  fit''
parameter selection may be quite challenging.  The process of choosing
these  parameters  generally  involves   the  comparison  of  a  given
observational data  set with the corresponding model  outputs.  Due to
the complexity of galaxy formation models, and the non-linear coupling
between  physical   prescriptions  in  the  model,   it  is  typically
non-trivial  to predict  how  variations of  parameters  or groups  of
parameters can  affect a given  model output.  We have  addressed this
problem by  implementing a sensitivity analysis,  which decomposes the
relationship  between   input  parameters  and   predicted  observable
properties into different  ``effects.''  Each effect characterizes how
an output responds to variations of only a subset of input parameters,
and thus can  be used to inform the user of  which parameters are most
important  and  most  likely  to  affect the  prediction  of  a  given
observable.  Conversely, this sensitivity analysis can also be used to
show what  model parameters can  be most efficiently constrained  by a
given  observational  data  set.   Finally, this  analysis  can  allow
modelers  to  simplify their  models,  or  at  the very  least  ignore
specific parameters for the purposes  of a given study, by identifying
input  parameters   that  have  no   affect  on  the   outputs  (i.e.,
observational predictions) of interest.

When applying a sensitivity analysis it is necessary to densely sample
one's model over the whole range of interest of its input parameter
space.  With the ChemTreeN code, doing this rapidly becomes
computationally prohibitive as the dimensionality of the input
parameter space increases (with each model requiring between
minutes and hours to complete).  To circumvent this problem, we have
trained statistical model emulators based on Gaussian processes.
These emulators act as an approximate (but reasonably accurate)
representation of the outputs of the ChemTreeN code and are very
computationally efficient, running in substantially less than a
millisecond, as opposed to hours for a ChemTreeN model.  This makes it
feasible to predict vast numbers of model outputs in a short period of
time.  While the Gaussian process emulator is an approximation of the
outputs from ChemTreeN, it is reasonably accurate when constructed
correctly \citep[as discussed in][]{G12}), and provides both an estimate
of the output values and their error (which is useful when comparing
the models to observational data).

We have  also shown how the  results of a sensitivity  analysis can be
easily  visualized  thanks to  the  ANOVA  decomposition.  The  ANOVA
decomposition  provides a way  to \textit{quantitatively}  measure the
percentage  of  the  total  output  parameter  variance  that  can  be
explained by  either variations  of single input  variables or  by any
combination of two or more input variables.

In the analysis performed in  this paper, we have considered the Milky
Way satellite galaxy  cumulative luminosity and metallicity functions.
Our work shows that, when the  luminosity function is used as the sole
observational  data set, the  parameters that  could be  most strongly
constrained  in  ChemTreeN are:  {\it  i}) the  onset  of  the epoch  of
reionization,  $z_{\rm  r}$,  {\it  ii}) the  baryonic  mass  fraction
assigned to  each dark matter  halo, i.e.  the baryon  accretion rate,
$f_{\rm  bary}$, and {\it  iii}) to  a much  lesser extent,  the star
formation efficiency,  $\epsilon_{*}$.  The decomposition  also showed
that the  bright end of the  luminosity function is  most sensitive to
baryon accretion  rate (and is essentially  \textit{only} sensitive to
this quantity), and the faint end  of the luminosity function is dominated
by variation in  the onset redshift of the epoch  of reionization.  We
also find through the ANOVA decomposition that the luminosity function
is  entirely insensitive to  variations of  the input  parameters that
control the remaining physical prescriptions.

Considering the Milky Way satellite galaxy metallicity function as the
observable  data  set allows  us  to  put  useful constraints  on  two
additional parameters:  the efficiency  with which metals  are ejected
from galaxies,  $f_{\rm esc}$, and  the efficiency of the  coupling of
supernova explosions with the interstellar medium within a galaxy (and
thus   their   energy    deposition   rates),   $\epsilon_{\rm   SN}$.
Interestingly,  we  find that  these  two  prescriptions are  strongly
non-linearly coupled.   The cumulative number of  metal-rich satellite
galaxies (defined here to be $\langle$[Fe/H]$\rangle   \geq  -1.1$)  is  strongly
dominated by the fraction of metals ejected out of the host galaxy due
to supernova-driven winds.   There is also  a weaker dependence  on the
efficiency of the supernova  energy coupling, $\epsilon_{\rm SN}$, and
on the redshift of reionization, $z_{\rm r}$.  As we move
toward lower values of $\langle$[Fe/H]$\rangle$ the cumulative number
of satellite  galaxies becomes strongly  sensitive to the  redshift of
the epoch of reionization.

It is important to remark that the ANOVA decomposition allows us to
characterize the relationship between the input parameters and the
desired model outputs independent of the corresponding \textit{real
observable values} and the underlying formation history of our
galactic model.  In other words, the relative relationship between the
model's input and output parameters is approximately independent of,
and thus separable from, the underlying galaxy merger history -- a
very useful result that can help to inform choices regarding the
quantity and properties of relatively expensive cosmological
simulations.  As a consequence of our ANOVA analysis, we were able to
reduce the dimensionality of the model's input parameter space from
seven dimensions to the five dimensions that actually had an impact on
the selected observables, and then to apply this
reduced-dimensionality model to our suite of cosmological simulations.

By defining  a statistical measure  of plausibility, and  by comparing
model emulators to mock observational  data in a quantitative manner, we have
demonstrated that it is possible to recover the input parameter vector
used to  create the  mock observational data  set \emph{even  when no
  prior knowledge of  the input parameter is provided}.  The search of
the  best-fitting  parameter  volume  required the  use  of a  Delayed
Rejection Adaptive Metropolis (DRAM)  method to sample the whole input
parameter  space. The involved  likelihood function  was based  on the
(im)plausibility  measure  defined  in  Section~\ref{sec:mod_emu}.   A
different choice  of implausibility measure (i.e.,  different means of
determining goodness-of-fit  of the model and  observational data) may
affect our results  in a quantitative sense, but is  unlikely to make
a qualitative difference in results for the observational quantities that we have chosen.

 We have applied  this statistical machinery to real observational
  data sets associated with the  Milky Way -- namely, the luminosity and
  metallicity  functions  of our  galaxy's  dwarf  satellites.  As  we
  showed in  G12, the best-fitting input  parameter selection process
  strongly depends on the  underlying merger history of Milky Way-like
  galaxy used to train the  model emulators. In this work we discussed
  how this characteristic of our method could be used to constrain the
  assembly history of  the Milky Way and its  properties at $z=0$. Our
  approach consisted of obtaining a  best fitting model for each Milky
  Way-like dark matter halo  being considered. The best-fit parameters
  were allowed to freely vary from  model to model.  In the four cases
  considered, the  Milky Way satellite luminosity  function allowed us
  to put  strong constraints not  only on the baryon  accretion rates,
  but also on how long the  less massive galaxies were able to accrete
  gas before reionization shut them down.  The best-fitting parameters
  showed   significant  scatter   between   cosmological  simulations,
  especially  along  the direction  of  the  baryon fraction,  $f_{\rm
    bary}$.   The  resulting  best-fit  models provided  a  luminosity
  function  that  tightly  fit   their  observed  counterpart  in  all
  cases. However,  only one  of our models  was able to  reproduce the
  observed stellar  halo mass  within 40 kpc  of the  Galactic center,
  $M_{40}$.  The remaining three  models showed values that are either
  substantially too  large or too  small when compared to  the observed
  value.  On  the basis of this  analysis it is  possible to disregard
  these three models and  their corresponding merger histories as good
  representations of  the underlying merger history of  the Milky Way.
  Interestingly, as previously observed by \citet{2010MNRAS.402.1995M}
  and  \citet{2013MNRAS.429..725S}, the  number of  luminous satellite
  galaxies  brighter  than  $M_{\rm  v}  =  -5$  shows  a  significant
  correlation with the host's dark matter halo virial mass.  Our preferred
  best-fitting  model, MW4, which  can reproduce the number  of bright
  satellites  with $M_{\rm v}  \leq -5$  and simultaneously  provide a
  reasonable estimate of $M_{40}$, has  a value of $M_{\rm vir} = 1.44
  \times 10^{12}$  $M_{\odot}$. This is in good  agreement with recent
  estimates   of  the   total   MW  mass   \citep{2010ApJ...720L.108G,
    2013ApJ...768..140B,   2013ApJ...764..161K,  2013arXiv1309.4293P}.
  Lower mass  halos such  as MW3,  with  $M_{\rm  vir} =  1.22 \times
  10^{12}$ $M_{\odot}$, would be ruled  out within our framework.  It
is  important  to notice  however  that  due  to the  relatively  poor
constraints obtained on the star formation efficiency, $\epsilon_{*}$,
two of the  models with different formation histories,  namely MW1 and
MW2,  presented multiple  likely  values for  this  parameter.  As  an
example,  for these  two  models  we computed  the  value of  $M_{40}$
associated with  the largest  plausible value of  $\epsilon_{*}$.  Our
results showed that, while one of the models was not able to reproduce
the observed value of $M_{40}$  even in this case, the second resulted
in  a much better  fit to  the observed  value.  Nonetheless,  in both
cases  the  mock  luminosity   functions  obtained  with  the  largest
plausible  value of $\epsilon_{*}$  resulted in  poorer fits  to their
observed  counterpart  than  those  obtained  with  the  best  fitting
parameters. A more robust  comparison could be achieved by contrasting
the  observed  $M_{40}$  to   the  distribution  of  modeled  $M_{40}$
associated with the  10 per cent most likely  values of $f_{\rm bary}$
and  $z_{\rm  r}$ and  their  corresponding  $\epsilon_{*}$.  This  is
beyond the scope of this paper and we defer it to a future work.

Due to incompleteness in our  current sample of satellite galaxies, as
well   as  uncertainties  on   $\langle$[Fe/H]$\rangle$  measurements,
results  based  on the  metallicity  function  are significantly  less
certain.  Incompleteness is  a much greater problem for  low mass (and
thus low  luminosity and low  metallicity) satellites; so,  to compare
with our  models we derive  a metallicity function that  only includes
galaxies  that  are  more  metal-rich than  $\langle$[Fe/H]$\rangle  =
-1.5$.  This  imposes a  magnitude limit of  M$_{\rm v}  \approx -11$,
which is within the luminosity  range of the classical dwarf galaxies.
In the  space defined  by $f_{\rm esc}$  and $\epsilon_{\rm  SN}$, the
DRAM sampling  of the  implausibility hyper surface associated with this
reduced  set  of galaxies  resulted  in  several  ``islands'' of  high
plausibility.  The models associated  with the best fitting parameters
presented,  in  all   cases,  a  luminosity-metallicity  relation  for
satellite galaxies  that agrees well  with the observed  relation over
that magnitude range.  However,  due to large source uncertainties, no
further constraints were obtained with this observational data set.

The results presented on this work are an example of a procedure that
can be applied to statistically constrain the formation history of the
Milky Way.  A more robust and statistically significant analysis would
require {\it i}) the addition of a large set of possible formation
histories and {\it ii}) a direct comparison to a larger number of
available observable quantities.  To address the first point, we are
currently running a large suite of high resolution dark matter-only
cosmological simulations of the formation of Milky Way-like halos.
These simulation will allow us to probe different galaxy formation
histories, ranging from halos that acquire most of their mass very
early on to halos that have had their last major merger episode close
to $z = 0$, and also the plausible range of masses that have been
attributed to the Milky Way.  The resulting best-fitting models
associated with each formation history will be confronted by a much
richer observational data set, including observable quantities such as
mean halo metallicity and chemical abundances as a function of radius,
radial distribution of satellite galaxies, and possibly even the
degree of phase-space substructure.  By using this iterative method,
we hope to provide useful constraints on the Milky Way's mass and
formation history that are complementary to alternate theoretical
techniques, and which provide insight both into our own Galaxy's
behavior and, more generally, into the process by which all galaxies
form.

\acknowledgments FAG, BWO, CCS and RLW are supported through the NSF
Office of Cyberinfrastructure by grant PHY-0941373.  FAG and BWO are
supported in part by the Michigan State University Institute for
Cyber-Enabled Research (iCER).  BWO was supported in part by the
Department of Energy through the Los Alamos National Laboratory
Institute for Geophysics and Planetary Physics and by NSF grant PHY
08-22648: Physics Frontiers Center/Joint Institute for Nuclear
Astrophysics (JINA). RLW is also supported by in part by NSF grant
DMS--0757549 and by NASA grant NNX09AK60G.

\bibliographystyle{apj}
\bibliography{apj-jour,halo_sa_emu}

\begin{thebibliography}{87}
\expandafter\ifx\csname natexlab\endcsname\relax\def\natexlab#1{#1}\fi

\bibitem[{{Barden} {et~al.}(2010){Barden}, {Jones}, {Barnes}, {Heijmans},
  {Heng}, {Knight}, {Orr}, {Smith}, {Churilov}, {Brzeski}, {Waller},
  {Shortridge}, {Horton}, {Mayfield}, {Haynes}, {Haynes}, {Whittard},
  {Goodwin}, {Smedley}, {Saunders}, {Gillingham}, {Penny}, {Farrell}, {Vuong},
  {Heald}, {Lee}, {Muller}, {Freeman}, {Bland-Hawthorn}, {Zucker}, \& {de
  Silva}}]{hermes}
{Barden}, S.~C., {et~al.} 2010, in Society of Photo-Optical Instrumentation
  Engineers (SPIE) Conference Series, Vol. 7735, Society of Photo-Optical
  Instrumentation Engineers (SPIE) Conference Series

\bibitem[{{Bell} {et~al.}(2008){Bell}, {Zucker}, {Belokurov}, {Sharma},
  {Johnston}, {Bullock}, {Hogg}, {Jahnke}, {de Jong}, {Beers}, {Evans},
  {Grebel}, {Ivezi{\'c}}, {Koposov}, {Rix}, {Schneider}, {Steinmetz}, \&
  {Zolotov}}]{bell08}
{Bell}, E.~F., {et~al.} 2008, \apj, 680, 295

\bibitem[{{Belokurov} {et~al.}(2006){Belokurov}, {Zucker}, {Evans}, {Gilmore},
  {Vidrih}, {Bramich}, {Newberg}, {Wyse}, {Irwin}, {Fellhauer}, {Hewett},
  {Walton}, {Wilkinson}, {Cole}, {Yanny}, {Rockosi}, {Beers}, {Bell},
  {Brinkmann}, {Ivezi{\'c}}, \& {Lupton}}]{2006ApJ...642L.137B}
{Belokurov}, V., {et~al.} 2006, \apjl, 642, L137

\bibitem[{{Belokurov} {et~al.}(2007){Belokurov}, {Zucker}, {Evans}, {Kleyna},
  {Koposov}, {Hodgkin}, {Irwin}, {Gilmore}, {Wilkinson}, {Fellhauer},
  {Bramich}, {Hewett}, {Vidrih}, {De Jong}, {Smith}, {Rix}, {Bell}, {Wyse},
  {Newberg}, {Mayeur}, {Yanny}, {Rockosi}, {Gnedin}, {Schneider}, {Beers},
  {Barentine}, {Brewington}, {Brinkmann}, {Harvanek}, {Kleinman}, {Krzesinski},
  {Long}, {Nitta}, \& {Snedden}}]{2007ApJ...654..897B}
---. 2007, \apj, 654, 897

\bibitem[{{Belokurov} {et~al.}(2010){Belokurov}, {Walker}, {Evans}, {Gilmore},
  {Irwin}, {Just}, {Koposov}, {Mateo}, {Olszewski}, {Watkins}, \&
  {Wyrzykowski}}]{2010ApJ...712L.103B}
---. 2010, \apjl, 712, L103

\bibitem[{{Benson}(2010)}]{2010PhR...495...33B}
{Benson}, A.~J. 2010, \physrep, 495, 33

\bibitem[{{Benson}(2012)}]{2012NewA...17..175B}
---. 2012, NewA, 17, 175

\bibitem[{{Bond} {et~al.}(2010){Bond}, {Ivezi{\'c}}, {Sesar}, {Juri{\'c}},
  {Munn}, {Kowalski}, {Loebman}, {Ro{\v s}kar}, {Beers}, {Dalcanton},
  {Rockosi}, {Yanny}, {Newberg}, {Allende Prieto}, {Wilhelm}, {Lee},
  {Sivarani}, {Majewski}, {Norris}, {Bailer-Jones}, {Re Fiorentin}, {Schlegel},
  {Uomoto}, {Lupton}, {Knapp}, {Gunn}, {Covey}, {Allyn Smith}, {Miknaitis},
  {Doi}, {Tanaka}, {Fukugita}, {Kent}, {Finkbeiner}, {Quinn}, {Hawley},
  {Anderson}, {Kiuchi}, {Chen}, {Bushong}, {Sohi}, {Haggard}, {Kimball},
  {McGurk}, {Barentine}, {Brewington}, {Harvanek}, {Kleinman}, {Krzesinski},
  {Long}, {Nitta}, {Snedden}, {Lee}, {Pier}, {Harris}, {Brinkmann}, \&
  {Schneider}}]{2010ApJ...716....1B}
{Bond}, N.~A., {et~al.} 2010, \apj, 716, 1

\bibitem[{{Bower} {et~al.}(2006){Bower}, {Benson}, {Malbon}, {Helly}, {Frenk},
  {Baugh}, {Cole}, \& {Lacey}}]{2006MNRAS.370..645B}
{Bower}, R.~G., {Benson}, A.~J., {Malbon}, R., {Helly}, J.~C., {Frenk}, C.~S.,
  {Baugh}, C.~M., {Cole}, S., \& {Lacey}, C.~G. 2006, \mnras, 370, 645

\bibitem[{{Bower} {et~al.}(2010){Bower}, {Vernon}, {Goldstein}, {Benson},
  {Lacey}, {Baugh}, {Cole}, \& {Frenk}}]{2010MNRAS.407.2017B}
{Bower}, R.~G., {Vernon}, I., {Goldstein}, M., {Benson}, A.~J., {Lacey}, C.~G.,
  {Baugh}, C.~M., {Cole}, S., \& {Frenk}, C.~S. 2010, \mnras, 407, 2017

\bibitem[{{Boylan-Kolchin} {et~al.}(2013){Boylan-Kolchin}, {Bullock}, {Sohn},
  {Besla}, \& {van der Marel}}]{2013ApJ...768..140B}
{Boylan-Kolchin}, M., {Bullock}, J.~S., {Sohn}, S.~T., {Besla}, G., \& {van der
  Marel}, R.~P. 2013, \apj, 768, 140

\bibitem[{{Bullock} \& {Johnston}(2005)}]{bj05}
{Bullock}, J.~S., \& {Johnston}, K.~V. 2005, \apj, 635, 931

\bibitem[{{Bullock} {et~al.}(2000){Bullock}, {Kravtsov}, \&
  {Weinberg}}]{2000ApJ...539..517B}
{Bullock}, J.~S., {Kravtsov}, A.~V., \& {Weinberg}, D.~H. 2000, \apj, 539, 517

\bibitem[{{Carollo} {et~al.}(2008){Carollo}, {Beers}, {Lee}, {Chiba}, {Norris},
  {Wilhelm}, {Sivarani}, {Marsteller}, {Munn}, {Bailer-Jones}, {Fiorentin}, \&
  {York}}]{2008Natur.451..216C}
{Carollo}, D., {et~al.} 2008, \nat, 451, 216

\bibitem[{{Carollo} {et~al.}(2010){Carollo}, {Beers}, {Chiba}, {Norris},
  {Freeman}, {Lee}, {Ivezi{\'c}}, {Rockosi}, \& {Yanny}}]{2010ApJ...712..692C}
---. 2010, \apj, 712, 692

\bibitem[{{Cooper} {et~al.}(2010){Cooper}, {Cole}, {Frenk}, {White}, {Helly},
  {Benson}, {De Lucia}, {Helmi}, {Jenkins}, {Navarro}, {Springel}, \&
  {Wang}}]{cooper}
{Cooper}, A.~P., {et~al.} 2010, \mnras, 406, 744

\bibitem[{{Corlies} {et~al.}(2013){Corlies}, {Johnston}, {Tumlinson}, \&
  {Bryan}}]{2013ApJ...773..105C}
{Corlies}, L., {Johnston}, K.~V., {Tumlinson}, J., \& {Bryan}, G. 2013, \apj,
  773, 105

\bibitem[{{Cui} {et~al.}(2012){Cui}, {Zhao}, {Chu}, {Li}, {Li}, {Zhang}, {Su},
  {Yao}, {Wang}, {Xing}, {Li}, {Zhu}, {Wang}, {Gu}, {Luo}, {Xu}, {Zhang},
  {Liu}, {Zhang}, {Yang}, {Cao}, {Chen}, {Chen}, {Chen}, {Chen}, {Chu}, {Feng},
  {Gong}, {Hou}, {Hu}, {Hu}, {Hu}, {Jia}, {Jiang}, {Jiang}, {Jiang}, {Jin},
  {Li}, {Li}, {Li}, {Liu}, {Liu}, {Lu}, {Mao}, {Men}, {Qi}, {Qi}, {Shi},
  {Tang}, {Tao}, {Wang}, {Wang}, {Wang}, {Wang}, {Wang}, {Wang}, {Wang},
  {Wang}, {Wang}, {Wang}, {Wang}, {Wang}, {Xu}, {Xu}, {Yang}, {Yu}, {Yuan},
  {Yuan}, {Zhai}, {Zhang}, {Zhang}, {Zhang}, {Zhao}, {Zhou}, {Zhou}, {Zhu}, \&
  {Zou}}]{lamost}
{Cui}, X.-Q., {et~al.} 2012, Research in Astronomy and Astrophysics, 12, 1197

\bibitem[{{Dekel} \& {Woo}(2003)}]{2003MNRAS.344.1131D}
{Dekel}, A., \& {Woo}, J. 2003, \mnras, 344, 1131

\bibitem[{{Diemand} {et~al.}(2006){Diemand}, {Kuhlen}, \& {Madau}}]{fof}
{Diemand}, J., {Kuhlen}, M., \& {Madau}, P. 2006, \apj, 649, 1

\bibitem[{{Font} {et~al.}(2006){Font}, {Johnston}, {Bullock}, \&
  {Robertson}}]{2006ApJ...638..585F}
{Font}, A.~S., {Johnston}, K.~V., {Bullock}, J.~S., \& {Robertson}, B.~E. 2006,
  \apj, 638, 585

\bibitem[{{Font} {et~al.}(2011){Font}, {McCarthy}, {Crain}, {Theuns}, {Schaye},
  {Wiersma}, \& {Dalla Vecchia}}]{2011MNRAS.416.2802F}
{Font}, A.~S., {McCarthy}, I.~G., {Crain}, R.~A., {Theuns}, T., {Schaye}, J.,
  {Wiersma}, R.~P.~C., \& {Dalla Vecchia}, C. 2011, \mnras, 416, 2802

\bibitem[{{Frebel} \& {Norris}(2013)}]{2013pss5.book...55F}
{Frebel}, A., \& {Norris}, J.~E. 2013, {Metal-Poor Stars and the Chemical
  Enrichment of the Universe}, ed. T.~D. {Oswalt} \& G.~{Gilmore}, 55

\bibitem[{Geweke(1992)}]{geweke}
Geweke, J. 1992, in IN BAYESIAN STATISTICS (University Press), 169--193

\bibitem[{{Gilbert} {et~al.}(2012){Gilbert}, {Guhathakurta}, {Beaton},
  {Bullock}, {Geha}, {Kalirai}, {Kirby}, {Majewski}, {Ostheimer}, {Patterson},
  {Tollerud}, {Tanaka}, \& {Chiba}}]{2012ApJ...760...76G}
{Gilbert}, K.~M., {et~al.} 2012, \apj, 760, 76

\bibitem[{{Gilmore} {et~al.}(2012){Gilmore}, {Randich}, {Asplund}, {Binney},
  {Bonifacio}, {Drew}, {Feltzing}, {Ferguson}, {Jeffries}, {Micela},
  {Negueruela}, {Prusti}, {Rix}, {Vallenari}, {Alfaro}, {Allende-Prieto},
  {Babusiaux}, {Bensby}, {Blomme}, {Bragaglia}, {Flaccomio}, {Fran{\c c}ois},
  {Irwin}, {Koposov}, {Korn}, {Lanzafame}, {Pancino}, {Paunzen},
  {Recio-Blanco}, {Sacco}, {Smiljanic}, {Van Eck}, \&
  {Walton}}]{2012Msngr.147...25G}
{Gilmore}, G., {et~al.} 2012, The Messenger, 147, 25

\bibitem[{{Girardi} {et~al.}(2002){Girardi}, {Bertelli}, {Bressan}, {Chiosi},
  {Groenewegen}, {Marigo}, {Salasnich}, \& {Weiss}}]{girardi2002}
{Girardi}, L., {Bertelli}, G., {Bressan}, A., {Chiosi}, C., {Groenewegen},
  M.~A.~T., {Marigo}, P., {Salasnich}, B., \& {Weiss}, A. 2002, \aap, 391, 195

\bibitem[{{Girardi} {et~al.}(2004){Girardi}, {Grebel}, {Odenkirchen}, \&
  {Chiosi}}]{girardi2004}
{Girardi}, L., {Grebel}, E.~K., {Odenkirchen}, M., \& {Chiosi}, C. 2004, \aap,
  422, 205

\bibitem[{{Gnedin}(2000)}]{gne}
{Gnedin}, N.~Y. 2000, \apj, 542, 535

\bibitem[{{Gnedin} {et~al.}(2010){Gnedin}, {Brown}, {Geller}, \&
  {Kenyon}}]{2010ApJ...720L.108G}
{Gnedin}, O.~Y., {Brown}, W.~R., {Geller}, M.~J., \& {Kenyon}, S.~J. 2010,
  \apjl, 720, L108

\bibitem[{{G{\'o}mez} {et~al.}(2012{\natexlab{a}}){G{\'o}mez}, {Coleman-Smith},
  {O'Shea}, {Tumlinson}, \& {Wolpert}}]{G12}
{G{\'o}mez}, F.~A., {Coleman-Smith}, C.~E., {O'Shea}, B.~W., {Tumlinson}, J.,
  \& {Wolpert}, R.~L. 2012{\natexlab{a}}, \apj, 760, 112

\bibitem[{{G{\'o}mez} {et~al.}(2013){G{\'o}mez}, {Helmi}, {Cooper}, {Frenk},
  {Navarro}, \& {White}}]{2013arXiv1307.0008G}
{G{\'o}mez}, F.~A., {Helmi}, A., {Cooper}, A.~P., {Frenk}, C.~S., {Navarro},
  J.~F., \& {White}, S.~D.~M. 2013, ArXiv e-prints

\bibitem[{{G{\'o}mez} {et~al.}(2012{\natexlab{b}}){G{\'o}mez}, {Minchev},
  {O'Shea}, {Lee}, {Beers}, {An}, {Bullock}, {Purcell}, \&
  {Villalobos}}]{2012MNRAS.423.3727G}
{G{\'o}mez}, F.~A., {et~al.} 2012{\natexlab{b}}, \mnras, 423, 3727

\bibitem[{{Greggio} \& {Renzini}(1983)}]{gregren}
{Greggio}, L., \& {Renzini}, A. 1983, \aap, 118, 217

\bibitem[{{Haario} {et~al.}(2006){Haario}, {Laine}, {Mira}, \&
  {Saksman}}]{dram}
{Haario}, H., {Laine}, M., {Mira}, A., \& {Saksman}, E. 2006, Statistics and
  Computing, 16, 339

\bibitem[{{Henriques} {et~al.}(2009){Henriques}, {Thomas}, {Oliver}, \&
  {Roseboom}}]{2009MNRAS.396..535H}
{Henriques}, B.~M.~B., {Thomas}, P.~A., {Oliver}, S., \& {Roseboom}, I. 2009,
  \mnras, 396, 535

\bibitem[{{Ivezi{\'c}} {et~al.}(2008){Ivezi{\'c}}, {Sesar}, {Juri{\'c}},
  {Bond}, {Dalcanton}, {Rockosi}, {Yanny}, {Newberg}, {Beers}, {Allende
  Prieto}, {Wilhelm}, {Lee}, {Sivarani}, {Norris}, {Bailer-Jones}, {Re
  Fiorentin}, {Schlegel}, {Uomoto}, {Lupton}, {Knapp}, {Gunn}, {Covey},
  {Smith}, {Miknaitis}, {Doi}, {Tanaka}, {Fukugita}, {Kent}, {Finkbeiner},
  {Munn}, {Pier}, {Quinn}, {Hawley}, {Anderson}, {Kiuchi}, {Chen}, {Bushong},
  {Sohi}, {Haggard}, {Kimball}, {Barentine}, {Brewington}, {Harvanek},
  {Kleinman}, {Krzesinski}, {Long}, {Nitta}, {Snedden}, {Lee}, {Harris},
  {Brinkmann}, {Schneider}, \& {York}}]{2008ApJ...684..287I}
{Ivezi{\'c}}, {\v Z}., {et~al.} 2008, \apj, 684, 287

\bibitem[{{Juri{\'c}} {et~al.}(2008){Juri{\'c}}, {Ivezi{\'c}}, {Brooks},
  {Lupton}, {Schlegel}, {Finkbeiner}, {Padmanabhan}, {Bond}, {Sesar},
  {Rockosi}, {Knapp}, {Gunn}, {Sumi}, {Schneider}, {Barentine}, {Brewington},
  {Brinkmann}, {Fukugita}, {Harvanek}, {Kleinman}, {Krzesinski}, {Long},
  {Neilsen}, {Nitta}, {Snedden}, \& {York}}]{2008ApJ...673..864J}
{Juri{\'c}}, M., {et~al.} 2008, \apj, 673, 864

\bibitem[{{Kallivayalil} {et~al.}(2013){Kallivayalil}, {van der Marel},
  {Besla}, {Anderson}, \& {Alcock}}]{2013ApJ...764..161K}
{Kallivayalil}, N., {van der Marel}, R.~P., {Besla}, G., {Anderson}, J., \&
  {Alcock}, C. 2013, \apj, 764, 161

\bibitem[{{Keller} {et~al.}(2012){Keller}, {Skymapper Team}, \& {Aegis
  Team}}]{2012ASPC..458..409K}
{Keller}, S.~C., {Skymapper Team}, \& {Aegis Team}. 2012, in Astronomical
  Society of the Pacific Conference Series, Vol. 458, Galactic Archaeology:
  Near-Field Cosmology and the Formation of the Milky Way, ed. W.~{Aoki},
  M.~{Ishigaki}, T.~{Suda}, T.~{Tsujimoto}, \& N.~{Arimoto}, 409

\bibitem[{{Kennedy} \& {O'Hagan}(2000)}]{Kenn:OHag:2000}
{Kennedy}, M.~C., \& {O'Hagan}, A. 2000, Biometrika, 1

\bibitem[{{Kirby} {et~al.}(2011){Kirby}, {Lanfranchi}, {Simon}, {Cohen}, \&
  {Guhathakurta}}]{2011ApJ...727...78K}
{Kirby}, E.~N., {Lanfranchi}, G.~A., {Simon}, J.~D., {Cohen}, J.~G., \&
  {Guhathakurta}, P. 2011, \apj, 727, 78

\bibitem[{{Klypin} {et~al.}(1999){Klypin}, {Kravtsov}, {Valenzuela}, \&
  {Prada}}]{1999ApJ...522...82K}
{Klypin}, A., {Kravtsov}, A.~V., {Valenzuela}, O., \& {Prada}, F. 1999, \apj,
  522, 82

\bibitem[{{Koposov} {et~al.}(2008){Koposov}, {Belokurov}, {Evans}, {Hewett},
  {Irwin}, {Gilmore}, {Zucker}, {Rix}, {Fellhauer}, {Bell}, \&
  {Glushkova}}]{2008ApJ...686..279K}
{Koposov}, S., {et~al.} 2008, \apj, 686, 279

\bibitem[{{Kroupa}(2001)}]{kroupa2001}
{Kroupa}, P. 2001, \mnras, 322, 231

\bibitem[{{Link} \& {Eaton}(2012)}]{thinning}
{Link}, W., \& {Eaton}, M. 2012, Methods in Ecology and Evolution, 3, 112

\bibitem[{{Lu} {et~al.}(2012){Lu}, {Mo}, {Katz}, \&
  {Weinberg}}]{2012MNRAS.421.1779L}
{Lu}, Y., {Mo}, H.~J., {Katz}, N., \& {Weinberg}, M.~D. 2012, \mnras, 421, 1779

\bibitem[{{Lu} {et~al.}(2013){Lu}, {Mo}, {Lu}, {Katz}, \&
  {Weinberg}}]{2013arXiv1311.0047L}
{Lu}, Y., {Mo}, H.~J., {Lu}, Z., {Katz}, N., \& {Weinberg}, M.~D. 2013, ArXiv
  e-prints

\bibitem[{{Macci{\`o}} {et~al.}(2010){Macci{\`o}}, {Kang}, {Fontanot},
  {Somerville}, {Koposov}, \& {Monaco}}]{2010MNRAS.402.1995M}
{Macci{\`o}}, A.~V., {Kang}, X., {Fontanot}, F., {Somerville}, R.~S.,
  {Koposov}, S., \& {Monaco}, P. 2010, \mnras, 402, 1995

\bibitem[{{Majewski} {et~al.}(2010){Majewski}, {Wilson}, {Hearty}, {Schiavon},
  \& {Skrutskie}}]{apogee}
{Majewski}, S.~R., {Wilson}, J.~C., {Hearty}, F., {Schiavon}, R.~R., \&
  {Skrutskie}, M.~F. 2010, in IAU Symposium, Vol. 265, IAU Symposium, ed.
  K.~{Cunha}, M.~{Spite}, \& B.~{Barbuy}, 480--481

\bibitem[{{McConnachie}(2012)}]{2012AJ....144....4M}
{McConnachie}, A.~W. 2012, \aj, 144, 4

\bibitem[{{McConnachie} {et~al.}(2009){McConnachie}, {Irwin}, {Ibata},
  {Dubinski}, {Widrow}, {Martin}, {C{\^o}t{\'e}}, {Dotter}, {Navarro},
  {Ferguson}, {Puzia}, {Lewis}, {Babul}, {Barmby}, {Bienaym{\'e}}, {Chapman},
  {Cockcroft}, {Collins}, {Fardal}, {Harris}, {Huxor}, {Mackey},
  {Pe{\~n}arrubia}, {Rich}, {Richer}, {Siebert}, {Tanvir}, {Valls-Gabaud}, \&
  {Venn}}]{2009Natur.461...66M}
{McConnachie}, A.~W., {et~al.} 2009, \nat, 461, 66

\bibitem[{{Monachesi} {et~al.}(2013){Monachesi}, {Bell}, {Radburn-Smith},
  {Vlaji{\'c}}, {de Jong}, {Bailin}, {Dalcanton}, {Holwerda}, \&
  {Streich}}]{anto}
{Monachesi}, A., {et~al.} 2013, \apj, 766, 106

\bibitem[{{Moore} {et~al.}(1999){Moore}, {Ghigna}, {Governato}, {Lake},
  {Quinn}, {Stadel}, \& {Tozzi}}]{1999ApJ...524L..19M}
{Moore}, B., {Ghigna}, S., {Governato}, F., {Lake}, G., {Quinn}, T., {Stadel},
  J., \& {Tozzi}, P. 1999, \apjl, 524, L19

\bibitem[{{Mouhcine} {et~al.}(2005){Mouhcine}, {Rich}, {Ferguson}, {Brown}, \&
  {Smith}}]{2005ApJ...633..828M}
{Mouhcine}, M., {Rich}, R.~M., {Ferguson}, H.~C., {Brown}, T.~M., \& {Smith},
  T.~E. 2005, \apj, 633, 828

\bibitem[{{Nomoto} {et~al.}(1997){Nomoto}, {Iwamoto}, {Nakasato}, {Thielemann},
  {Brachwitz}, {Tsujimoto}, {Kubo}, \& {Kishimoto}}]{nomo1997}
{Nomoto}, K., {Iwamoto}, K., {Nakasato}, N., {Thielemann}, F.-K., {Brachwitz},
  F., {Tsujimoto}, T., {Kubo}, Y., \& {Kishimoto}, N. 1997, Nuclear Physics A,
  621, 467

\bibitem[{{Oakley} \& {O'Hagan}(2002)}]{Oakl:Ohag:2002}
{Oakley}, J.~E., \& {O'Hagan}, A. 2002, Biometrika, 89, 769

\bibitem[{{Oakley} \& {O'Hagan}(2004)}]{Oakl:Ohag:2004}
---. 2004, Journal of the Royal Statistical Society: Series B (Statistical
  Methodology), 66

\bibitem[{O'Hagan(2006)}]{OHag:2006}
O'Hagan, A. 2006, Reliability Engineering \& System Safety, 91, 1290 , the
  Fourth International Conference on Sensitivity Analysis of Model Output (SAMO
  2004) - SAMO 2004

\bibitem[{{Perryman} {et~al.}(2001){Perryman}, {de Boer}, {Gilmore}, {H{\o}g},
  {Lattanzi}, {Lindegren}, {Luri}, {Mignard}, {Pace}, \& {de
  Zeeuw}}]{2001A&A...369..339P}
{Perryman}, M.~A.~C., {et~al.} 2001, \aap, 369, 339

\bibitem[{{Piffl} {et~al.}(2013){Piffl}, {Scannapieco}, {Binney}, {Steinmetz},
  {Scholz}, {Williams}, {de Jong}, {Kordopatis}, {Matijevic}, {Bienayme},
  {Bland-Hawthorn}, {Boeche}, {Freeman}, {Gibson}, {Gilmore}, {Grebel},
  {Helmi}, {Munari}, {Navarro}, {Parker}, {Reid}, {Seabroke}, {Watson}, {Wyse},
  \& {Zwitter}}]{2013arXiv1309.4293P}
{Piffl}, T., {et~al.} 2013, ArXiv e-prints

\bibitem[{{Radburn-Smith} {et~al.}(2011){Radburn-Smith}, {de Jong}, {Seth},
  {Bailin}, {Bell}, {Brown}, {Bullock}, {Courteau}, {Dalcanton}, {Ferguson},
  {Goudfrooij}, {Holfeltz}, {Holwerda}, {Purcell}, {Sick}, {Streich}, {Vlajic},
  \& {Zucker}}]{2011ApJS..195...18R}
{Radburn-Smith}, D.~J., {et~al.} 2011, \apjs, 195, 18

\bibitem[{Rasmussen \& Williams(2005)}]{Rasmussen05}
Rasmussen, C.~E., \& Williams, C. K.~I. 2005, {Gaussian Processes for Machine
  Learning (Adaptive Computation and Machine Learning)} (The MIT Press)

\bibitem[{{Ruiz} {et~al.}(2013){Ruiz}, {Cora}, {Padilla}, {Dom{\'{\i}}nguez},
  {Tecce}, {Orsi}, {Yaryura}, {Garc{\'{\i}}a Lambas}, {Gargiulo}, \& {Mu{\~n}oz
  Arancibia}}]{2013arXiv1310.7034R}
{Ruiz}, A.~N., {et~al.} 2013, ArXiv e-prints

\bibitem[{{Sacks} {et~al.}(1989){Sacks}, {Welch}, {Mitchell}, \&
  {Wynn}}]{Sack:Welc:Mitc:Wynn:1989}
{Sacks}, J., {Welch}, W.~J., {Mitchell}, T.~J., \& {Wynn}, H.~P. 1989, Stat.\
  Sci., 4, 409

\bibitem[{{Santner} {et~al.}(2003){Santner}, {Williams}, \&
  {Notz}}]{Sant:Will:Notz:2003}
{Santner}, T.~J., {Williams}, B.~J., \& {Notz}, W. 2003, {The Design and
  Analysis of Computer Experiments}

\bibitem[{{Schonlau} \& {Welch}(2006)}]{X}
{Schonlau}, M., \& {Welch}, W.~J. 2006, Screening the Input Variables to a
  Computer Code Via Analysis of Variance and Visualization, ed. A.~{Dean} \&
  S.~{Lewis} (New York: Springer New York)

\bibitem[{{Siebert}(2012)}]{2012sf2a.conf..121S}
{Siebert}, A. 2012, in SF2A-2012: Proceedings of the Annual meeting of the
  French Society of Astronomy and Astrophysics, ed. S.~{Boissier}, P.~{de
  Laverny}, N.~{Nardetto}, R.~{Samadi}, D.~{Valls-Gabaud}, \& H.~{Wozniak},
  121--127

\bibitem[{{Smith} {et~al.}(2008){Smith}, {Szidarovszky}, {Karnavas}, \&
  {Bahill}}]{sa_1}
{Smith}, E.~D., {Szidarovszky}, F., {Karnavas}, W.~J., \& {Bahill}, A.~T. 2008,
  The Open Cybernetics \& Systemics Journal, 2, 39

\bibitem[{{Spergel} {et~al.}(2007){Spergel}, {Bean}, {Dor{\'e}}, {Nolta},
  {Bennett}, {Dunkley}, {Hinshaw}, {Jarosik}, {Komatsu}, {Page}, {Peiris},
  {Verde}, {Halpern}, {Hill}, {Kogut}, {Limon}, {Meyer}, {Odegard}, {Tucker},
  {Weiland}, {Wollack}, \& {Wright}}]{wmap}
{Spergel}, D.~N., {et~al.} 2007, \apjs, 170, 377

\bibitem[{{Springel}(2005)}]{springel2005}
{Springel}, V. 2005, \mnras, 364, 1105

\bibitem[{{Springel} {et~al.}(2008){Springel}, {Wang}, {Vogelsberger},
  {Ludlow}, {Jenkins}, {Helmi}, {Navarro}, {Frenk}, \&
  {White}}]{2008MNRAS.391.1685S}
{Springel}, V., {et~al.} 2008, \mnras, 391, 1685

\bibitem[{{Starkenburg} {et~al.}(2013){Starkenburg}, {Helmi}, {De Lucia}, {Li},
  {Navarro}, {Font}, {Frenk}, {Springel}, {Vera-Ciro}, \&
  {White}}]{2013MNRAS.429..725S}
{Starkenburg}, E., {et~al.} 2013, \mnras, 429, 725

\bibitem[{{Steinmetz} {et~al.}(2006){Steinmetz}, {Zwitter}, {Siebert},
  {Watson}, {Freeman}, {Munari}, {Campbell}, {Williams}, {Seabroke}, {Wyse},
  {Parker}, {Bienaym{\'e}}, {Roeser}, {Gibson}, {Gilmore}, {Grebel}, {Helmi},
  {Navarro}, {Burton}, {Cass}, {Dawe}, {Fiegert}, {Hartley}, {Russell},
  {Saunders}, {Enke}, {Bailin}, {Binney}, {Bland-Hawthorn}, {Boeche}, {Dehnen},
  {Eisenstein}, {Evans}, {Fiorucci}, {Fulbright}, {Gerhard}, {Jauregi}, {Kelz},
  {Mijovi{\'c}}, {Minchev}, {Parmentier}, {Pe{\~n}arrubia}, {Quillen}, {Read},
  {Ruchti}, {Scholz}, {Siviero}, {Smith}, {Sordo}, {Veltz}, {Vidrih}, {von
  Berlepsch}, {Boyle}, \& {Schilbach}}]{rave}
{Steinmetz}, M., {et~al.} 2006, \aj, 132, 1645

\bibitem[{{Tammann} {et~al.}(1994){Tammann}, {Loeffler}, \&
  {Schroeder}}]{1994ApJS...92..487T}
{Tammann}, G.~A., {Loeffler}, W., \& {Schroeder}, A. 1994, \apjs, 92, 487

\bibitem[{{Tissera} {et~al.}(2013{\natexlab{a}}){Tissera}, {Beers}, {Carollo},
  \& {Scannapieco}}]{2013arXiv1309.3609T}
{Tissera}, P., {Beers}, T., {Carollo}, D., \& {Scannapieco}, C.
  2013{\natexlab{a}}, ArXiv e-prints

\bibitem[{{Tissera} {et~al.}(2013{\natexlab{b}}){Tissera}, {Scannapieco},
  {Beers}, \& {Carollo}}]{2013MNRAS.432.3391T}
{Tissera}, P.~B., {Scannapieco}, C., {Beers}, T.~C., \& {Carollo}, D.
  2013{\natexlab{b}}, \mnras, 432, 3391

\bibitem[{{Tollerud} {et~al.}(2008){Tollerud}, {Bullock}, {Strigari}, \&
  {Willman}}]{toll08}
{Tollerud}, E.~J., {Bullock}, J.~S., {Strigari}, L.~E., \& {Willman}, B. 2008,
  \apj, 688, 277

\bibitem[{{Tominaga}(2009)}]{2009ApJ...690..526T}
{Tominaga}, N. 2009, \apj, 690, 526

\bibitem[{{Tumlinson}(2006)}]{tumlinson1}
{Tumlinson}, J. 2006, \apj, 641, 1

\bibitem[{{Tumlinson}(2010)}]{tumlinson2}
---. 2010, \apj, 708, 1398

\bibitem[{{Wang} {et~al.}(2013){Wang}, {Frenk}, \&
  {Cooper}}]{2013MNRAS.429.1502W}
{Wang}, J., {Frenk}, C.~S., \& {Cooper}, A.~P. 2013, \mnras, 429, 1502

\bibitem[{{Widrow} {et~al.}(2012){Widrow}, {Gardner}, {Yanny}, {Dodelson}, \&
  {Chen}}]{2012ApJ...750L..41W}
{Widrow}, L.~M., {Gardner}, S., {Yanny}, B., {Dodelson}, S., \& {Chen}, H.-Y.
  2012, \apjl, 750, L41

\bibitem[{{Williams} {et~al.}(2013){Williams}, {Steinmetz}, {Binney},
  {Siebert}, {Enke}, {Famaey}, {Minchev}, {de Jong}, {Boeche}, {Freeman},
  {Bienaym{\'e}}, {Bland-Hawthorn}, {Gibson}, {Gilmore}, {Grebel}, {Helmi},
  {Kordopatis}, {Munari}, {Navarro}, {Parker}, {Reid}, {Seabroke}, {Sharma},
  {Siviero}, {Watson}, {Wyse}, \& {Zwitter}}]{2013MNRAS.tmp.2356W}
{Williams}, M.~E.~K., {et~al.} 2013, \mnras

\bibitem[{{Willman} {et~al.}(2004){Willman}, {Governato}, {Dalcanton}, {Reed},
  \& {Quinn}}]{will04}
{Willman}, B., {Governato}, F., {Dalcanton}, J.~J., {Reed}, D., \& {Quinn}, T.
  2004, \mnras, 353, 639

\bibitem[{{Yanny} {et~al.}(2009){Yanny}, {Rockosi}, {Newberg}, {Knapp},
  {Adelman-McCarthy}, {Alcorn}, {Allam}, {Allende Prieto}, {An}, {Anderson},
  {Anderson}, {Bailer-Jones}, {Bastian}, {Beers}, {Bell}, {Belokurov},
  {Bizyaev}, {Blythe}, {Bochanski}, {Boroski}, {Brinchmann}, {Brinkmann},
  {Brewington}, {Carey}, {Cudworth}, {Evans}, {Evans}, {Gates}, {G{\"a}nsicke},
  {Gillespie}, {Gilmore}, {Nebot Gomez-Moran}, {Grebel}, {Greenwell}, {Gunn},
  {Jordan}, {Jordan}, {Harding}, {Harris}, {Hendry}, {Holder}, {Ivans},
  {Ivezi{\v c}}, {Jester}, {Johnson}, {Kent}, {Kleinman}, {Kniazev},
  {Krzesinski}, {Kron}, {Kuropatkin}, {Lebedeva}, {Lee}, {French Leger},
  {L{\'e}pine}, {Levine}, {Lin}, {Long}, {Loomis}, {Lupton}, {Malanushenko},
  {Malanushenko}, {Margon}, {Martinez-Delgado}, {McGehee}, {Monet}, {Morrison},
  {Munn}, {Neilsen}, {Nitta}, {Norris}, {Oravetz}, {Owen}, {Padmanabhan},
  {Pan}, {Peterson}, {Pier}, {Platson}, {Re Fiorentin}, {Richards}, {Rix},
  {Schlegel}, {Schneider}, {Schreiber}, {Schwope}, {Sibley}, {Simmons},
  {Snedden}, {Allyn Smith}, {Stark}, {Stauffer}, {Steinmetz}, {Stoughton},
  {SubbaRao}, {Szalay}, {Szkody}, {Thakar}, {Sivarani}, {Tucker}, {Uomoto},
  {Vanden Berk}, {Vidrih}, {Wadadekar}, {Watters}, {Wilhelm}, {Wyse}, {Yarger},
  \& {Zucker}}]{segue}
{Yanny}, B., {et~al.} 2009, \aj, 137, 4377

\bibitem[{{York} {et~al.}(2000){York}, {Adelman}, {Anderson}, {Anderson},
  {Annis}, {Bahcall}, {Bakken}, {Barkhouser}, {Bastian}, {Berman}, {Boroski},
  {Bracker}, {Briegel}, {Briggs}, {Brinkmann}, {Brunner}, {Burles}, {Carey},
  {Carr}, {Castander}, {Chen}, {Colestock}, {Connolly}, {Crocker}, {Csabai},
  {Czarapata}, {Davis}, {Doi}, {Dombeck}, {Eisenstein}, {Ellman}, {Elms},
  {Evans}, {Fan}, {Federwitz}, {Fiscelli}, {Friedman}, {Frieman}, {Fukugita},
  {Gillespie}, {Gunn}, {Gurbani}, {de Haas}, {Haldeman}, {Harris}, {Hayes},
  {Heckman}, {Hennessy}, {Hindsley}, {Holm}, {Holmgren}, {Huang}, {Hull},
  {Husby}, {Ichikawa}, {Ichikawa}, {Ivezi{\'c}}, {Kent}, {Kim}, {Kinney},
  {Klaene}, {Kleinman}, {Kleinman}, {Knapp}, {Korienek}, {Kron}, {Kunszt},
  {Lamb}, {Lee}, {Leger}, {Limmongkol}, {Lindenmeyer}, {Long}, {Loomis},
  {Loveday}, {Lucinio}, {Lupton}, {MacKinnon}, {Mannery}, {Mantsch}, {Margon},
  {McGehee}, {McKay}, {Meiksin}, {Merelli}, {Monet}, {Munn}, {Narayanan},
  {Nash}, {Neilsen}, {Neswold}, {Newberg}, {Nichol}, {Nicinski}, {Nonino},
  {Okada}, {Okamura}, {Ostriker}, {Owen}, {Pauls}, {Peoples}, {Peterson},
  {Petravick}, {Pier}, {Pope}, {Pordes}, {Prosapio}, {Rechenmacher}, {Quinn},
  {Richards}, {Richmond}, {Rivetta}, {Rockosi}, {Ruthmansdorfer}, {Sandford},
  {Schlegel}, {Schneider}, {Sekiguchi}, {Sergey}, {Shimasaku}, {Siegmund},
  {Smee}, {Smith}, {Snedden}, {Stone}, {Stoughton}, {Strauss}, {Stubbs},
  {SubbaRao}, {Szalay}, {Szapudi}, {Szokoly}, {Thakar}, {Tremonti}, {Tucker},
  {Uomoto}, {Vanden Berk}, {Vogeley}, {Waddell}, {Wang}, {Watanabe},
  {Weinberg}, {Yanny}, {Yasuda}, \& {SDSS Collaboration}}]{sdss}
{York}, D.~G., {et~al.} 2000, \aj, 120, 1579

\end{thebibliography}

\label{lastpage}
\end{document}